\documentclass[twocolumn,showpacs,
 amsmath,amssymb,
 aps,
pra,
]{revtex4-1}

\usepackage{graphicx}
\usepackage{dcolumn}
\usepackage{bm}
\usepackage{subeqnarray}
\usepackage{cases}

\begin{document}

\preprint{APS/123-QED}

\title{Terminal Velocity Motion Model Used to Analyze the Mutual Phase-locking of STNOs}

\author{Hao-Hsuan Chen}
\email{HaoHsuanChen@hotmail.com}
\affiliation{Hefei Innovation Research Institute, Beihang University, Hefei, 230013, China}

\author{Ching-Ming Lee}
\email{cmlee@yuntech.edu.tw}
\affiliation{Graduate School of Materials Science, National Yunlin University of Science and Technology, Douliou, 64002, Taiwan}

\author{Ching-Ray Chang}
\affiliation{%
Department of Physics, National Taiwan University, Taipei 10617, Taiwan
}%





\date{\today}

\begin{abstract}
Using Legendre transformation, a standard theoretical approach extensively used in classical mechanics as well as thermal dynamics, two-dimensional non-linear auto-oscillators including spin torque nano-oscillators (STNOs) can be equivalently expressed either in phase space or in configuration space where all of them can be modeled by terminal velocity motion (TVM) particles.  The transformation completely preserves the dynamic information about the canonical momenta, leading to very precise analytical predictions about the phase-locking of a coupled pair of perpendicular to plane STNOs (PERP-STNOs) including dynamical phase diagrams, (un)phase-locked  frequencies, phase-locked angles, and transient evolutions, which are all solved based on Newton mechanics. Notably, the TVM model successfully solves the difficulty of the generalized pendulum-like model [Chen \textit{et al}. \textbf{J. Appl. Phys. 130}, 043904 (2021)] failing to make precise predictions for the higher range of current in serial connection. Additionally, how to simply search for the critical currents for phase-locked (PL) and asynchronized (AS) states by numerically simulating the macrospin as well as TVM model, which gets inspired through analyzing the excitations of a forced pendulum, is also supplied here. Therefore, we believe that the TVM model can bring a more intuitive and precise way to explore all types of two-dimensional non-linear auto-oscillators.
\end{abstract}

\maketitle


\section{\label{sec:level1}INTRODUCTION}
Spin-Torque Nano-Oscillators (STNOs) driven by a well-known anti-damping effect induced by spin-polarization effect \cite{slonczewski1996current,berger1996emission,Slonczewski2002,XiaoJiang2004,Hirsch1999} or spin Hall effect (SHE) \cite{Demidov2014,Awad2017,Awad2018} have drawn a lot of attention due to them generating magnetic auto-oscillations in the GHz to sub-THz frequency range, giving rise to several promising applications such as microwave emitters\cite{kiselev2003microwave}, wireless communication\cite{Consolo2010,Choi2014}, as well as neuromorphic computation\cite{Romera2018}. There have so far been several reported types of STNOs, including ones based on the quasi-uniform mode in nano-pillars (NPs) \cite{kiselev2003microwave,Houssameddine2007,Kubota2013},  nano-contacts (NCs) \cite{kaka2005}, NC-spin Hall nano-oscillators (NC-SHNOs) \cite{Demidov2014,Awad2017,Awad2018}, non-uniform magnetic solitons \cite{Pribiag2007,Khvalkovskiy2009,Hoefer2010,XiaoD2017,Garcia-Sanchez2016}, and anti-ferromagnetism\cite{Cheng2016,Khymyn2017,Shen2019}.

However, there are still some tough issues in practical application of STNOs, such as low emitted power and large linewidth.
Up to now, the most frequently proposed and feasible idea to handle these issues has been to synchronize an array of multiple STNOs via some coupling mechanisms. Several types of coupling mechanisms have so far been reported, containing propagating spin waves based on NC structure \cite{kaka2005,Mancoff2005}; electric coupling in the circuit based on NP structure \cite{grollier2006,Taniguchi2018}; magnetic dipolar coupling based on NPs structure (quasi-uniform or vortex modes) \cite{HaoHsuan2011,HaoHsuan2012,Belanovsky2012,AbreuAraujo2015,HaoHsuan2016,Kang2018,HaoHsuan2018,Mancilla-Almonacid2019,Chen2021, Li2017}, NCs structure (droplets) \cite{Wang2017,Wang2018a}, and NC-SHNOs \cite{Awad2018,Zahedinejad2020}.

Previously we have adopted a generalized pendulum-like model developed based on the expansions at free-running states to analytically solve the initial condition (IC)-dependent excitations of mutually phase-locked (PL) as well as asynchronized (AS) states in a pair of perpendicular-to-plane STNOs (PERP-STNOs) coupled by a magnetic dipolar interaction\cite{Chen2021}, which certainly fails to be explained by the Kuramoto model\cite{Acebron2005}. Here, according to the pendulum-like model\cite{Chen2021}, the key point to cause these IC-dependent excitations is the existence of intrinsic kinetic-like energies of oscillators, which is induced by their non-linear frequency shift coefficients making their frequency change with their oscillating amplitude. IC-dependent excitations have actually been observed and analyzed in the pioneering works on the injection-locking of other types of STNOs\cite{Bonin2009,Tabor2010,Zhou2010,Li2010,DongLi2011,DAquino2017,Tortarolo2018}, manifesting that these phenomena should be the common character of all types of STNOs with considerable non-linear frequency shift coefficients and thereby can be well elucidated by the pendulum-like model.

Although the generalized pendulum-like model makes successful predictions in most cases for the synchronization of PERP-STNOs\cite{Chen2021}, there are still some shortages about this model failing to make precise predictions for the higher range of current in serial connection. The reason for that, as Ref.\cite{Chen2021} pointed out, is that the canonical momenta should be variables rather than the fixed values given from the free-running states of oscillators during the whole evolution of phase-locking. That is, once the final momenta of the phase-locked state are far from those of the free-running states, the predictions made by the pendulum-like model will deviate very much from the macrospin simulation results, implying that the dynamical information about momenta possessed by this model is inadequate. In reality, in the respective of classical mechanics, pendulum is a mechanical system whose equation of motion can be expressed either in configuration space or in phase space, which means that there has been a theoretical technique termed \textit{Legendre transformation}\cite{goldstein2014classical}, a standard theoretical approach extensively used in classical mechanics as well as thermal dynamics, to make the expression of its governing equation transfer between these two spaces without loss of the information about momenta or phase angle velocities. Conversely speaking, the Landau-Lifshitz-Gilbert-Slonczewski(LLGS) equation, which has been expressed in terms of canonical variables\cite{Chen2021}, can also be equivalently expressed in configuration space with certainty through this transformation.

In this paper, we aim to develop a new theoretical model based on Legendre transformation to intuitively analyze the mechanisms of IC-dependent mutual phase-locking in a coupled pair of PERP-STNOs. The paper is organized as follows: In section \ref{A}, we take the example of the injection-locking of a PERP-STNO to a circularly polarized oscillating field as a starting point to explain why Legendre transformation is useful in analyzing the phase-locking of non-linear oscillators and what can be benefited from expressing the governing equations of non-linear oscillators in configuration space. In section \ref{B} and Appendix \ref{appa} we develop a new theoretical framework, termed the terminal velocity motion (TVM) model, based on the Legendre transformation. Moreover, we take several examples of oscillators of various kinds including linear TVM particle, perpendicular magnetized anisotropy STNO (PMA-STNO), PERP-STNO, and ven der Pol oscillator (quasi-linear oscillator) to prove all of their dynamics can be well expressed as well as analyzed by the TVM model. In section \ref{C}, the analytical approaches based on the TVM model are developed to calculate the critical currents, the frequency responses for PL and AS states, respectively, and phase-locked phase angles in a coupled pair of PERP-STNOs by a magnetic dipolar coupling . In section \ref{D}, a simple operational procedure inspired by trigging a forced pendulum is offered to numerically solve the dynamical phase diagrams, the frequency responses and phase-locked phase angles for the coupled PERP-STNO pairs, which can be used to verify the analytical results. Finally, a brief summary and discussion about the benefits as well as perspects of the TVM model in analyses of mutual synchronization of non-linear oscillators compared with our previous theory are given in section \ref{sec:3}.

\section{\label{sec:2}Model and Theory}
\subsection{\label{A}Motivations of using TVM model to analyze mutual phase-locking of STNOs}
\begin{figure}
\begin{center}
\includegraphics[width=2.6cm]{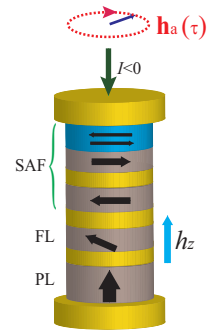}
\end{center}
\caption{Schematic of injection locking of a PERP-STNO induced by a circularly polarized oscillating field, where $\mathbf{h}_{\mathrm{a}}(t)=h_{\mathrm{a}}[\mathbf{\hat{x}}
\cos(\omega_{\mathrm{e}}t+\Psi)
+\mathbf{\hat{y}}\sin(\omega_{\mathrm{e}}t+\Psi)]$, $h_{z}$ is an applied field normal to the thin film plane, and P, F, and SAF indicate the pinned, free, and synthetic anti-ferromagnetic layers, respectively. $I$ is an
injected current.}
\label{DevStruc_circuH}
\end{figure}

To explain our motivations of developing a new theory for analyzing the mutual phase-locking of STNOs, one can begin with briefly presenting the analytical procedure of solving the phase-locking of a PERP-STNO to a weak external oscillating field, which is termed \textit{injection locking} and has been studied by the group of G. Bertotti et.al \cite{Bonin2009,GBertotti2009nonlinear}
. As depicted in Fig.\ref{DevStruc_circuH}, a PERP-STNO is composed of a spin polarizer layer (P) with a fixed magnetization normal to the thin film, a free layer (F) with an in-plane magnetization, and a synthetic antiferromagnetic (SAF) trilayer as an analyzer. Then, applying a circularly polarized oscillating external field with a fixed angular frequency $\omega_{\mathrm{e}}$ and a sufficiently small amplitude $h_{\mathrm{a}}$ to the STNO is used to generate the phase-locking of the oscillator. Besides, $h_{z}$ is an applied field perpendicular to the thin film plane, which can be used to change the range of the generated frequency of the oscillator.
In the following, the equation of motion governing the dynamics of the STNO can be obtained by projecting the scaled Landau-Lifshitz-Gilbert-Slonczewski(LLGS) equation onto a canonical coordinate system (or, phase space) $(p\equiv-m_{z},\phi)$
 (see Appendix \ref{appa:1} in for the derails):
\setlength\abovedisplayskip{6pt}
\setlength\belowdisplayskip{6pt}
\begin{eqnarray}
\dot{p}&=&-\alpha S(p)\dot{\phi}+\beta(p,\mu)-\frac{\partial H}{\partial\phi},\nonumber\\
\dot{\phi}&=&\frac{\partial H}{\partial p},
\label{injectlocking}
\end{eqnarray}
where the scaled total Hamiltonian $H$ contains two terms: one is the dominating energy term including the demagnetization energy and the zeeman energy that is induced from $h_{z}$, i.e. $H_{\mathrm{o}}(p)=(1/2)p^{2}+h_{z}p$; the other is the minor zeeman energy term of interaction with a weak oscillating applied field $\mathbf{h}_{\mathrm{a}}(t)$, i.e. $H_{\mathrm{I}}(p,\phi,t)=-\mathbf{h}_{\mathrm{a}}\cdot\mathbf{m}=-h_{\mathrm{a}}\sqrt{1-p^{2}}\cos(\phi-\omega_{\mathrm{e}}t-\Psi)$, where $\Psi$ is the initial phase angle of the oscillating field. The first and second terms on the right-hand side of the first sub-equation of Eq. (\ref{injectlocking}) are the positive damping effect with a Gilbert damping constant $\alpha$ and positive damping function $S(p)=(1-p^{2})$ and the negative damping effect given by the spin-transfer torque (STT) $\beta(p,\mu)=a_{J0}(1-p^2)(p\Lambda^{2})/[(\Lambda^{2}+1)+(\lambda^{2}-1)]$.

We have known that there are two criteria for stable auto-oscillations in the free-running regime of an auto-oscillator operation, where $H_{I}$ is absent \cite{Chen2021}: one is $\dot{p}=-\alpha S(p_{0})(\frac{\partial H}{\partial p})_{p_{0}}+\beta(p_{0},\mu)=0$, meaning that the positive and negative damping contributions to the time-averaged energy consumption have to exactly compensate to each other in order to have the magnetization lasting precessing on a certain orbit labeled by $p_{0}$; the other is $(\partial^{2}H_{\mathrm{o}}/\partial p^{2})_{p_{0}}>(\partial[\beta(p,\mu)/\alpha S(p)]/\partial p)_{p_{0}}$, ensuring that once small fluctuations in the momentum $p$ is present there must be a \textit{restoring force} pulling it back to its free-running value $p_{0}$, i.e. the stability of an STNO.

On the basis of the existence of free-running regime, in the phase-locked regime of an auto-oscillator operation, where $H_{I}$ is present, there are two criteria for equilibrium phase-locked states $(p_{0},\Phi_{0})$, which are very similar to solving ferromagnetic resonance (FMR) states:
\setlength\abovedisplayskip{6pt}
\setlength\belowdisplayskip{6pt}
\begin{eqnarray}
\dot{p}&=&-\alpha S(p_{0})\omega_{\mathrm{e}}+\beta(p_{0},\mu)-h_{\mathrm{a}}\sqrt{1-p_{0}^{2}}\sin(\Phi_{0}-\Psi)\nonumber\\
&=&0,\nonumber\\
\dot{\Phi}&=&\left[\omega(p_{0})-\omega_{\mathrm{e}}\right]+h_{\mathrm{a}}\left(\frac{p_{0}}{\sqrt{1-p_{0}^{2}}}\right)
\cos(\Phi_{0}-\Psi)=0,\nonumber\\
\label{injectlocking_creterion}
\end{eqnarray}
where $\Phi\equiv\phi-\omega_{\mathrm{e}}t$ and $\omega(p)=\partial H_{\mathrm{o}}/\partial p=p+h_{z}$. Here, the first criterion requires that the contributions of the positive, negative damping, and oscillating external field to the time-averaged energy dissipation have to achieve a new balance to keep a permanent oscillation of the magnetization; the second one requires that the STNO has to be synchronous to the oscillating field, i.e. $\dot{\phi}=\omega_{\mathrm{e}}$. To confirm the stability of the equilibrium points $(p_{0},\Phi_{0})$, we further require the following criteria:
\setlength\abovedisplayskip{6pt}
\setlength\belowdisplayskip{6pt}
\begin{eqnarray}
\mathrm{det}(\mathbf{C}_{0})&>&0,\nonumber\\
\mathrm{Tr}(\mathbf{C}_{0})&<&0,\nonumber
\label{}
\end{eqnarray}
 where the matrix $\mathbf{C}_{0}$ is given by the linear expansion of Eq. (\ref{injectlocking}) around the equilibrium orientations of the magnetization $(p_{0},\Phi_{0})$:
\setlength\abovedisplayskip{6pt}
\setlength\belowdisplayskip{6pt}
\begin{eqnarray}
\frac{d}{dt}\left(
\begin{array}{l}
\delta p\\
\delta\Phi
\end{array}
\right)=\mathbf{C}_{0}(p_{0},\Phi_{0})\left(
\begin{array}{l}
\delta p\\
\delta\Phi
\end{array}
\right).\nonumber
\label{}
\end{eqnarray}

Although it is easy to obtain the exact solutions of the stable phase-locked states in the case of injection-locking following the above procedure, it would be hard to generalize that procedure to other cases more complicated like mutual phase-locking among multiple STNOs. Due to the fact of the sufficient smallness of the external oscillating field, where in the absence of the non-conservative effect the conservative field responsible for the precession of the STNO is dominating, i.e. $|\omega(p)|=|\partial H_{\mathrm{o}}/\partial p|\gg h_{\mathrm{a}}$, the above second requirement $\dot{\Phi}=0$ can be well simplified to be $\omega(p_{0})\approx\omega_{\mathrm{e}}$, that is, the generated frequency of the STNO has to be close enough to the frequency of the oscillating field.

 Thus, for a linear oscillator whose natural frequency doesn't depend on the momentum $p$, i.e. $\omega(p)=\omega_{0}=\mathrm{constant}$ or, for an example, the demagnetization field is absent in our presented system ($\omega(p)=h_{z}$), the frequency of the external driving source $\omega_{\mathrm{e}}$ has to \textit{actively} approach $\omega_{0}$ in order to generate the phase-locking. However, in contrast with a linear oscillator, for a non-linear oscillator with a momentum-dependent natural frequency, just as with the presence of the demagnetization field in our presented system ($\omega(p)=p+h_{z}$), its generated frequency $\omega(p)$ has to \textit{actively} approach $\omega_{\mathrm{e}}$ instead through adjusting the momentum $p$ to realize the phase-locking. That means that, strictly speaking, the phase-locked momentum $p_{0}$ should be decided by the condition $\omega(p_{0})\approx\omega_{\mathrm{e}}$ instead of solving from the free-running regime requiring $-\alpha S(p_{0})\dot{\phi}+\beta(p_{0},\mu)\approx0$, which is the result of the assumption $h_{\mathrm{a}}\ll|\beta|$ and $\alpha$ \cite{Slavin2009,Chen2021}. Once the solved $p_{0}$ is far from its free-running state value to break the original balance made by the positive/negative damping effects, then the extra energy consumption has to be offset by the oscillating field whose energy injection rate can be adjusted by the relative phase angle $\Phi-\Psi$ between the oscillating field and the STNO to ensure $\dot{p}=0$, that is, the phase-locked angle $\Phi_{0}$ can be solved by the first criterion in Eq. (\ref{injectlocking_creterion}). Thus, the criteria for the stability of $p_{0},\Phi_{0}$ will be simplified as follows: one is $(\partial^{2}H_{\mathrm{o}}/\partial p^{2})_{p_{0}}>(\partial[\beta'(p,\Phi_{0},\mu)/\alpha S(p)]/\partial p)_{p_{0}}$\cite{Chen2021}, where the effective negative damping function is $\beta'(p,\Phi_{0},\mu)\equiv\beta(p,\mu)-h_{\mathrm{a}}\sqrt{1-p^{2}}\sin(\Phi_{0}-\Psi)$; the other is $(\partial^{2} H_{I}/\partial\Phi^{2})_{(p_{0},\Phi_{0})}>0$.

Based on the smallness of $h_{\mathrm{a}}$, an analytical technique can be provided to obtain the phase-locked $p_{0}$ as a function of $\omega_{\mathrm{e}}$:
\setlength\abovedisplayskip{6pt}
\setlength\belowdisplayskip{6pt}
\begin{eqnarray}
p_{0}&=&\omega^{-1}\left[\omega_{\mathrm{e}}-h_{\mathrm{a}}f(p_{0})
\cos(\Phi_{0}-\Psi)\right],\nonumber\\
&=&\omega^{-1}[\omega_{\mathrm{e}}-h_{\mathrm{a}}f(\omega^{-1}\left[\omega_{\mathrm{e}}-h_{\mathrm{a}}f(p_{0})
\cos(\Phi_{0}-\Psi)\right])\nonumber\\
&&\times\cos(\Phi_{0}-\Psi)],\nonumber\\
&=&\omega^{-1}(\omega_{\mathrm{e}})-\left(\frac{\partial\omega^{-1}}{\partial X}\right)_{X=\omega_{\mathrm{e}}}h_{\mathrm{a}}\nonumber\\
&&\times f(\omega^{-1}\left[\omega_{\mathrm{e}}-h_{\mathrm{a}}f(p_{0})\cos(\Phi_{0}-\Psi)\right])\cos(\Phi_{0}-\Psi)\nonumber\\
&&+O(h_{\mathrm{a}}^{2}),\nonumber\\
&\approx&\omega^{-1}(\omega_{\mathrm{e}})-\left(\frac{\partial\omega}{\partial p}\right)^{-1}_{p=\omega^{-1}(\omega_{\mathrm{e}})}h_{\mathrm{a}}f(\omega^{-1}(\omega_{\mathrm{e}}))\nonumber\\
&&\times\cos(\Phi_{0}-\Psi),\nonumber
\label{}
\end{eqnarray}
where $f(p_{0})=p_{0}/\sqrt{1-p^{2}_{0}}$, and where $p_{0}$ appearing in the oscillating field has been repeatedly replaced by
$\omega^{-1}\left[\omega_{\mathrm{e}}-h_{\mathrm{a}}f(p_{0})\cos(\Phi_{0}-\Psi)\right]$. Then, the phase-locked $\Phi_{0}$ can be obtained from the equality
\setlength\abovedisplayskip{6pt}
\setlength\belowdisplayskip{6pt}
\begin{eqnarray}
&&-\alpha S(\omega^{-1}(\omega_{\mathrm{e}}))\left(\frac{\partial H}{\partial p}\right)_{\omega^{-1}(\omega_{\mathrm{e}})}+\beta(\omega^{-1}(\omega_{\mathrm{e}}),\mu)\nonumber\\
&&-h_{\mathrm{a}}\sqrt{1-\left[\omega^{-1}(\omega_{\mathrm{e}})\right]^{2}}
\sin(\Phi_{0}-\Psi)\approx0,\nonumber
\label{}
\end{eqnarray}
where the terms related to the factors $\alpha h_{\mathrm{a}}$, $|(\partial\beta/\partial p)_{p=\omega^{-1}(\omega_{\mathrm{e}})}|h_{\mathrm{a}}$, $h^{2}_{\mathrm{a}}$ are all much smaller than $\alpha$, $|\beta|$, and $h_{\mathrm{a}}$ to be neglected.
In reality, we can also express $p$ as a function of $\dot{\phi}$ in advance in a similar way:
\setlength\abovedisplayskip{6pt}
\setlength\belowdisplayskip{6pt}
\begin{eqnarray}
p&\approx&\omega^{-1}(\dot{\phi})-\left(\frac{\partial\omega}{\partial p}\right)^{-1}_{p=\omega^{-1}(\dot{\phi})}h_{\mathrm{a}}f(\omega^{-1}(\dot{\phi}))\nonumber\\
&&\times\cos(\Phi-\Psi),
\label{pasafucphi}
\end{eqnarray}
then implementing $\dot{\phi}=\omega_{\mathrm{e}}$ to Eq. (\ref{pasafucphi}) can complete the same procedure given previously. Notably, Eq. (\ref{pasafucphi}) can completely be transformed back into the second sub-equation of Eq. (\ref{injectlocking}) using the similar technique like what has been done in expressing $p$ as a function of $\dot{\phi}$, indicating that this replacement Eq. (\ref{pasafucphi}) is reversible.

Interestingly, Eq. (\ref{pasafucphi}) gives quite an inspiring hint that the LLGS equation expressed in \textit{phase space} $(p,\phi)$, i.e. Eq. (\ref{injectlocking}), can be turned into an equivalent expressed in \textit{configuration space} $(\dot{\phi},\phi)$ through a particular sort of transformation, which is termed \textit{Legendre transformation} and has been extensively used in classical mechanics and thermodynamics. According to the variational principle introduced in classical mechanics\cite{goldstein2014classical}, since the angular velocity $\dot{\phi}$, in contrast with the momentum $p$, is not an another independent variable to $\phi$, which means that once $\phi$ makes a change like $\phi\rightarrow\phi+\omega_{\mathrm{e}}t$ then $\dot{\phi}$ must have a corresponding change $\dot{\phi}\rightarrow\dot{\phi}+\omega_{\mathrm{e}}$, then it is more straightforwardly to solve the phase-locked state by requiring $\dot{\phi}=\omega_{\mathrm{e}}$ in configuration space than by solving $p_{0}$ in advance in phase space.

More importantly, in configuration space the phase-locking of a non-linear auto-oscillator can be well understood in a more intuitive way, i.e. a Newtonian particle, than in phase space, and, undoubtedly, the accuracy of the new theory developed in the following has been actually foreseen by Eq. (\ref{pasafucphi}) and Legendre transformation. Therefore, based on these motivations, we would like to generalize Eq. (\ref{pasafucphi}) to the case of an array of multiple non-linear aut-oscillators with weak interactions to obtain a new theory for analyzing the phase-locking of a coupled pair PERP-STNOs.

\subsection{\label{B}Generalized Terminal Velocity Motion}
For a group of two-dimensional autonomous particles with weak interactions, the equation of motion can be expressed in a general vector form as
\setlength\abovedisplayskip{6pt}
\setlength\belowdisplayskip{6pt}
\begin{eqnarray}
\frac{d\mathbf{x}_{i}}{dt}
&=&-(\nabla_{\mathbf{x}_{i}} E)\times\mathbf{e}_{n}-\alpha(\mathbf{x}_{i})
\frac{d\mathbf{x}_{i}}{dt}\nonumber\\
&&+\beta_{i}(\mathbf{x}_{i},\mu_{i})\left[(\mathbf{e}_{n}\times\mathbf{e}_{p})\times\mathbf{e}_{n}\right],
\label{vectformappro}
\end{eqnarray}
where $\mathbf{x}_{i}$ describes the state of the particles on a two
dimensional phase plane (see Ref.\cite{Chen2021}). The terms on the right-hand side of Eq. (\ref{vectformappro}) are in turns conservative, positive and negative dissipation parts, respectively. Also, $E$ is the total energy which includes non-linear intrinsic dynamic energies $E_{0i}$ of individual particles and weak interactive potential energirs $U_{I}$. $\alpha_{i}$ and $\beta_{i}$ are positive and negative factors, respectively. Expressing Eq. (\ref{vectformappro}) in terms of a generalized cyclic coordinate, i.e.
\textit{energy-phase angle} representation and using the \textit{Legendre} transformation, one obtains the approximated equivalent of Eq. (\ref{vectformappro}) in the configuration space $(\phi_{i},\dot{\phi}_{i})$ (see Appendix \ref{appa}):
\setlength\abovedisplayskip{6pt}
\setlength\belowdisplayskip{6pt}
\begin{eqnarray}
\ddot{\phi}_{i}&=&\left[\frac{1}{m_{\mathrm{eff},i}(\omega_{oi}^{-1}(\dot{\phi}_{i}))}\right]
\Bigg[-\alpha_{i}S_{i}(\omega_{oi}^{-1}(\dot{\phi}_{i}))\dot{\phi}_{i}
\nonumber\\
&&+\beta_{i}(\omega_{oi}^{-1}(\dot{\phi}_{i}),\mu_{i})-\frac{\partial H_{I}}{\partial\phi_{i}}\Bigg]+A\sum_{l=1}^{n}\frac{\partial q_{i}}{\partial\phi_{l}}\dot{\phi}_{l},
\label{nonconserNewtonb}
\end{eqnarray}
where $m_{\mathrm{eff},i}(\omega_{oi}^{-1}(\dot{\phi}_{i}))\equiv(d\omega_{oi}/dp_{i})_{\omega_{oi}^{-1}(\dot{\phi}_{i})}^{-1}$ is the effective mass, and $Aq_{i}(\dot{\phi}_{j},\phi_{j})=(\partial H_{I}/\partial p_{i})_{p_{j}=\omega^{-1}_{oj}(\dot{\phi}_{j})}$, where $A$ is the strength of interactions. $U_{I}$, $S_{i}$, and $\beta_{i}$ are the interaction energy, positive, and negative damping functions, respectively.

\begin{figure*}
\begin{center}
\includegraphics[width=18cm]{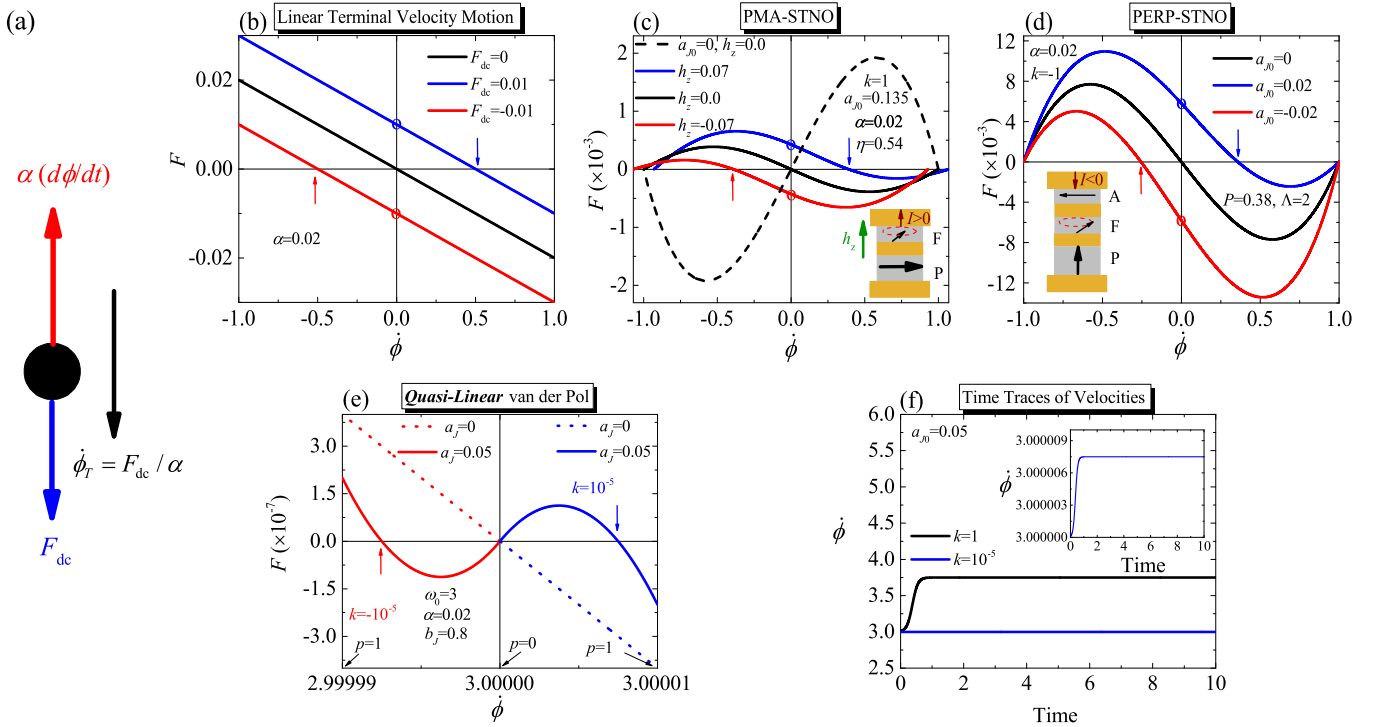}
\end{center}
\caption{(Color Online) Generalization of a linear TVM model to STNOs and quasi-linear oscillators.  (a) Schematic of a linear TVM particle. The red and blue arrows indicate the damping force $\alpha\dot{\phi}$ and the constant driving force $F_{\mathrm{dc}}$, respectively. The black arrow indicates the terminal velocity $\dot{\phi}_{T}=F_{\mathrm{dc}}/\alpha$. (b)Linear TVM model. The damping constant $\alpha$ is taken to be $0.02$. The black, blue, and red straight lines correspond to $F_{\mathrm{dc}}=0,0.01$, and $-0.01$, respectively. The blue and red arrows mark the terminal velocities for $F_{\mathrm{dc}}=0.01$, and $-0.01$, respectively. The blue and red 'O' mark the positions where the values of $F_{\mathrm{dc}}$ are taken. (c) and (d) are the TVM models for PMA-STNO and PERP-STNO, respectively, where the damping constants for them are both set to be $0.02$, the spin-polarization factors for these two sorts of STNOs are $\eta=0.54$ and $(P,\Lambda)=(0.38,2)$, respectively, and the anisotropic constants $k$ for them are taken as $1$ and $-1$, respectively. The panels in (c) and (d) are schematics of PMA-STNOs and PERP-STNOs, respectively, where P, F, and A indicate the pinned, free, and analyzer layers, respectively. Also, the green and brown arrows indicate the external magnetic field normal to the film plane and current injection direction, respectively.  In (c), the dashed curve indicates the case of the absence of current and field. The black, blue, and red solid curves denote the cases of $h_{z}=0,0.07$, and $-0.07$, respectively, which are under the same injection of $a_{J0}=0.135$. In (d), the black, blue, and red solid solid curves denote the cases of $a_{J0}=0,0.02$, and $-0.02$, respectively. (e) is the TVM model for a quasi-linear van der Pol oscillator, where the red ($k=-10^{-5}$) and blue ($k=10^{-5}$) frequency shift cases are plotted within $\dot{\phi}\in(3-10^{-5},\omega_{0}=3)$ and $\dot{\phi}\in(\omega_{0}=3,3+10^{-5})$, respectively, where the range of the momentum is $p\in(0,1)$ , and where $\alpha=0.02$, $b_{J}=0.8$, and the solid and dashed curves indicate $a_{J}=0.05$ and $a_{J}=0$, respectively. (f)Transient time evolution of $\dot{\phi}$ of the van der Pol oscillator. The black and blue solid curves indicate $k=1$ and $10^{-5}$, respectively. The panel is the zoom in for the case of $k=10^{-5}$.   }
\label{termivelomotion}
\end{figure*}

 For an individual non-linear oscillator in the absence of $U_{I}=0$, Eq. (\ref{nonconserNewtonb}) becomes
\setlength\abovedisplayskip{6pt}
\setlength\belowdisplayskip{6pt}
\begin{eqnarray}
\ddot{\phi_{i}}&=&\left[\frac{1}{m_{\mathrm{eff},i}(\omega_{oi}^{-1}(\dot{\phi}_{i}))}\right]\big[-\alpha_{i}S_{i}(\omega_{oi}^{-1}(\dot{\phi}_{i}))
\dot{\phi}_{i}\nonumber\\
&&+\beta_{i}(\omega_{oi}^{-1}(\dot{\phi}_{i}),\mu_{i})\big].
\label{nonconserNewtonc}
\end{eqnarray}
 Before taking insight into the dynamics of a non-linear free-running \textit{auto}-oscillator using Eq. (\ref{nonconserNewtonb}), one can take a first look at its simplest form, i.e. a linear \textit{terminal velocity motion} (TVM) particle (see Fig. \ref{termivelomotion}(a)), whose equation of motion can be written as
\setlength\abovedisplayskip{6pt}
\setlength\belowdisplayskip{6pt}
\begin{eqnarray}
\ddot{\phi}&=&F_{\mathrm{dc}}-\alpha\dot{\phi},
\label{lineartermimotion}
\end{eqnarray}
where $F_{\mathrm{dc}}$ is a time constant negative damping force, and the effective mass has been normalized to unity. Although it has a linear appearance, it is indeed non-linear in terms of phase space $(p,\phi)$ (see Eq. (\ref{apppphi})): through the Legendre transformation its Hamiltonian is $H_{O}(p)=(1/2)p^{2}$ from its Lagrangian $L(\dot{\phi})=(1/2)\dot{\phi}^{2}$. Then, we have $\dot{\phi}=\omega_{o}(p)=p$, and its positive/damping factors are both constants in phase space, i.e. $S(p)=1$ and $\beta(p,\mu)=F_{\mathrm{dc}}$. Thus, according to the criteria of a stable \textit{limit cycle} in phase space \cite{Chen2021}, we easily get the cycle solution labeled by the momentum $p$: $p_{0}=F_{\mathrm{dc}}/\alpha$ and its stability confirmation: $\left(\partial^{2} H_{O}/\partial p^{2}\right)_{p_{0}}=1>\left(\partial (F_{\mathrm{dc}}/\alpha)/\partial p\right)_{p_{0}}=0$. By the way, this kind of motion can also be expressed in terms of the \textit{Universal model}, a well-known theoretical framework for auto-oscillators, which is introduced in Ref. \cite{Slavin2009}. Here, if we take the pair of canonical variables $(|p|,\phi)$ as the \textit{power} $p$ ($p>0$) and \textit{phase} $\phi$, respectively, defined in that model, we get the angular frequency $\omega(p)$ and the positive/negative damping rates as follows:
\setlength\abovedisplayskip{6pt}
\setlength\belowdisplayskip{6pt}
\begin{eqnarray}
\omega(p)&=&|p|,\nonumber\\
\Gamma_{+}(p)&=&\frac{\alpha S(|p|)\left(\frac{\partial H_{O}}{\partial p}\right)}{2|p|}=\frac{\alpha}{2},\nonumber\\
\Gamma_{-}(p)&=&\frac{\beta(|p|,\mu)}{2|p|}=\frac{|F_{\mathrm{dc}}|}{2|p|},\nonumber
\label{}
\end{eqnarray}
respectively, then the cycle solution labeled by the power $p$ along with its stability confirmation are obtained as follows: $p_{0}=|F_{\mathrm{dc}}|/\alpha>0$ and $\left(\partial\Gamma_{+}/\partial p\right)_{p_{0}}=0>\left(\partial\Gamma_{-}/\partial p\right)_{p_{0}}=-(|F_{\mathrm{dc}}|/\alpha)/p^{2}$, respectively.

Eq. (\ref{lineartermimotion}) concludes that there are four basic ingredients to form a terminal velocity motion: \textbf{a}. effective mass of inertia $(m_{\mathrm{eff}}=1)$, which is related to the non-linearity of the dynamic state energy; \textbf{b}. positive damping force $(-\alpha\dot{\phi},\alpha>0)$; \textbf{c}. constant negative (anti-)damping force $F_{\mathrm{dc}}$, whose existence can be confirmed at $\dot{\phi}=0$, as marked by "o" symbols in Fig. \ref{termivelomotion}(b);
\textbf{d}. stability of terminal velocity $\dot{\phi}_{T}=F_{\mathrm{dc}}/\alpha$, which can be given as follows: Putting $\dot{\phi}(t)=\dot{\phi}_{T}+\delta\dot{\phi}$ into Eq. (\ref{lineartermimotion}), we have $\delta\ddot{\phi}=-\alpha\delta\dot{\phi}=\kappa\delta\dot{\phi}$ with $\kappa=-\alpha<0$, implying that any small deviation
in $\dot{\phi}$ that is away from $\dot{\phi}_{T}$ will be dragged back to $\dot{\phi}_{T}$ by the velocity-restoring force. This can also be seen quite straightforwardly from the negative slope of the force $F=F_{\mathrm{dc}}-\alpha\dot{\phi}$ against $\dot{\phi}$, which is taken at the crossing points with $F=0$, as indicated by the red/blue arrows in Fig. \ref{termivelomotion}(b). We would like to stress here that for the third ingredient, there are two equivalent ways to generate $F_{\mathrm{dc}}$ (see Fig. \ref{termivelomotion}(b)): one is to \textit{actively} add a constant force $F_{\mathrm{dc}}$ to $F$; the other is to \textit{passively} give the particle a velocity boosting, i.e. $\dot{\phi}\rightarrow\dot{\phi}+\omega$, making the positive damping force $-\alpha\dot{\phi}$ produce an extra constant force $F_{\mathrm{dc}}=-\alpha\omega$. These two equivalent perspectives for producing $F_{\mathrm{dc}}$ will give us a very intuitive and interesting explantation about why some kinds of STNOs need an external field to assist their auto-oscillations, which will be discussed in the following.

Having concluded the ingredients for a stable TVM motion previously,  one can rewrite Eq. (\ref{nonconserNewtonc}) as
$\ddot{\phi_{i}}=F_{i}(\dot{\phi}_{i})$, where $F_{i}(\dot{\phi}_{i})\equiv[1/m_{\mathrm{eff},i}(\dot{\phi}_{i})][f_{\mathrm{p}i}(\dot{\phi}_{i})+f_{\mathrm{n}i}(\dot{\phi}_{i})]$, and then the criteria for the stable equilibrium dynamic states $\dot{\phi}_{i0}(\neq 0)$ are demanding $F_{i}(\dot{\phi}_{i0})=0$ as well as $F_{i}^{(1)}(\dot{\phi}_{i0})=[\partial F_{i}(\dot{\phi}_{i})/\partial\dot{\phi}_{i}]_{\dot{\phi}_{i0}}<0$. Here, $f_{\mathrm{p(n)}i}(\dot{\phi}_{i})$ are the positive/negative damping forces, respectively.

Using the generalized TVM model (Eq. (\ref{nonconserNewtonb})), we would like to briefly analyze the dynamics of the two types of STNOs, namely PERP-STNOs and PMA-STNOs in a more intuitive way and point out the differences between them. In terms of the cylindrical coordinate system $(m_{z},\phi)$, their dynamic state energies can be written as $E_{0}=(-k/2)m^{2}_{z}$, where $k>0$ belongs to PMA-STNOs and $k<0$ corresponds to PERP-STNOs. If $m_{z}$ and $E_{0}$  are taken as the canonical momentum $p\equiv-m_{z}$ and  Hamiltonian $H_{O}$ (see \cite{Chen2021}), respectively, then the effective masses $m_{\mathrm{eff}}(\dot{\phi})=-k^{-1}$ for them will have opposite signs to each other, meaning that the angular acceleration $\ddot{\phi}$ in PMA-STNOs will be anti-parallel to the force $F$ (red shift); while in PERP-STNOs it will be parallel to $F$ (blue shift). Interestingly, in such a perspective, the impact of the damping force $f_{\mathrm{p}}(\dot{\phi})=-\alpha S(p)\dot{\phi}=-\alpha S(\dot{\phi})\dot{\phi}=-\alpha(1-\dot{\phi}^{2})\dot{\phi}$ (see Appendix \ref{appa:2} for the details) on a PMA-STNO will be equivalent to a \textit{negative} damping one due to its negative mass ($m_{\mathrm{eff}}=-1$), where the effective particle with a small deviation away from $\dot{\phi}=0$ will go through speeding up to $\dot{\phi}=\pm1$, as illustrated by the dashed curve in Fig. \ref{termivelomotion}(c). Here, $k$ has been normalized to be unity. In contrast, the impact of the damping force on a PERP-STNO is still positive damping, as reflected by the black curve in Fig. \ref{termivelomotion}(d).

Inspired by those four gradients about a TVM particle, we know PMA-STNOs and PERP-STNOs need a \textit{positive} damping force and a negative damping one, respectively, to compensate their intrinsic negative and positive damping ones, respectively, which can be achieved only by injecting an STT. In PMA-STNO, the \textit{positive} damping force has the form to be $f_{\mathrm{n}}(\dot{\phi})=-a_{J0}(-p)[1-(1-\lambda^{2}(1-p^{2}))^{-1/2}]=-a_{J0}\dot{\phi}[1-(1-\lambda^{2}
(1-\dot{\phi}^{2}))^{-1/2}]$\cite{Taniguchi_2013}, where $a_{J0}=\hbar\gamma\eta I/2eM_{s}V$ is the STT strength and $a_{J0}$ has to be positive here otherwise $f_{\mathrm{n}}(\dot{\phi})$ will enhance $f_{\mathrm{p}}(\dot{\phi})$ instead. $\lambda=\eta^{2}$ and $\eta$ are spin-polarization factors controlling the asymmetry and strength of the STT. Notably, without the asymmetric factors appearing in the STT with an in-plane spin-polarization (see the panel of Fig. \ref{termivelomotion}(c)), the STT fails to be an effective damping-like force \cite{Taniguchi_2013}. Furthermore, it should be noticed that the presence of $f_{\mathrm{n}}(\dot{\phi})$ can only turn the intrinsic \textit{negative} $f_{\mathrm{p}}(\dot{\phi})$ at most into a \textit{positive} damping force to give the PMA-STNO a potential stability, where there is still a lack of a dc negative damping force $F_{\mathrm{dc}}$ at $\dot{\phi}=0$ to drive an auto-oscillation, as can be seen by the solid black curve in Fig. \ref{termivelomotion}(c).

Obviously, according to the previous discussion about how to generate $F_{\mathrm{dc}}$, here we only have to give the system a velocity boosting by applying an external field $h_{z}$. Adding a zeeman energy to the PMA-STNO's Hamiltonian  $H_{O}(p)=-(1/2)p^{2}+h_{z}p$, we get $\dot{\phi}=-p+h_{z}$ ($\dot{\phi}\in(-1+h_{z},1+h_{z})$) where the velocity $\dot{\phi}$ has been boosted by $h_{z}$ that has been normalized by $k=1$. Then, replacing $p$ with $h_{z}-\dot{\phi}$, we have $f_{\mathrm{p}}(\dot{\phi})=-\alpha\dot{\phi}[1-(\dot{\phi}-h_{z})^{2}]$ and $f_{\mathrm{n}}(\dot{\phi})=-a_{J0}(\dot{\phi}-h_{z})\{1-[1-\lambda^{2}(1-(\dot{\phi}-h_{z})^{2})]^{-1/2}\}$. Thereby, a dc force $F_{\mathrm{dc}}=a_{J0}h_{z}\{1-[1-\lambda^{2}(1-h_{z}^{2})]^{-1/2}\}$ is obtained from $f_{\mathrm{n}}(\dot{\phi}=0)$, which confirms the existence of a TVM particle, i.e. a stable auto-oscillation, as indicated by the blue/red arrows in Fig. \ref{termivelomotion} (c).

In PERP-STNOs, the negative damping is $f_{\mathrm{n}}(\dot{\phi})=a_{J0}(1-\dot{\phi}^{2})(P\Lambda^{2})/[(\Lambda^{2}+1)+(\Lambda^{2}-1)(-\dot{\phi})]$, which has a dc negative damping force $F_{\mathrm{dc}}=(a_{J0}P\Lambda^{2})/(\Lambda^{2}+1)$ and thereby confirms the existence of a stable auto-oscillation without the assistance of an applied field, as can be seen from Fig. \ref{termivelomotion}(d). Additionally, we
would like to stress here that the STT that exists in PERP-STNOs only serves as a dc negative damping force rather than as a positive damping one, thus it is impossible to offer stability to PMA-STNO by this STT. On the contrary, the STT existing in PMA-STNO will be turned into a positive/negative damping in PERP-STNO with $a_{J0}<0(>0)$, which only enhances/weakens the stability of PERP-STNOs but fails to create a dc negative damping force, that is, either an STT with a PERP polarizer or an external field $h_{z}$ is still needed here to trigger an oscillation.

As a small summary, in the perspective of the TVM model we get good qualitative and quantitative analyses about why PMA-STNOs need an applied field to assist their auto-oscillations, but PERP-STNOs don't, which have been well verified by the numerical simulations\cite{Ebels2008,JHChang2011,HaoHsuan2011,HaoHsuan2015,HaoHsuan2017,Chen2019,Chen2021,Taniguchi_2013}
and experiments\cite{Houssameddine2007,Kubota2013}.

Finally, we would like to stress that TVM model can completely cover the case of \textit{quasi-linear} auto-oscillators with an extremely small non-linear frequency shift coefficient. We here take a well-known example called \textit{van der Pol} oscillator (see Ref. \cite{Slavin2009} for all the details) to understand quasi-linear auto-oscillators in the perspective of the TVM model: The Hamiltonian is $H_{O}=\omega_{0}p+(k/2)p^{2}$, then we have $\dot{\phi}=\omega_{0}+kp$, where the constant frequency $\omega_{0}$ in an LC-circuit is given by $\omega_0=1/\sqrt{LC}$ and $k$ is the non-linear frequency shift coefficient. For a quasi-linear oscillator, $k$ is an extremely small number $(|k|\ll1)$ that makes an exceedingly small range of $\dot{\phi}$ within a finite range of the momentum $p\in(0,1)$, that is, $\dot{\phi}\in(\omega_{0},\omega_{0}+kp)$ for $(k>0)$ or $\dot{\phi}\in(\omega_{0}+kp,\omega_{0})$ for $(k<0)$, and a huge effective mass $m_{\mathrm{eff}}=k^{-1}$. According to the example taken in Ref. \cite{Slavin2009}, where the equation of motion of a van der Pol oscillator is expressed by the universal model, we get the positive/negative damping rates $\Gamma_{+}(p)=\alpha$ and $\Gamma_{+}(p)=a_{J}(1-b_{J}p)$, respectively, where $\alpha\equiv R/2L$, $aJ\equiv S_{0}/2C$, and $b_{J}\equiv\alpha_{2}/4$\cite{Slavin2009}. If we take the \textit{power} $p$ and \textit{phase angle} $\phi$  defined in Ref. \cite{Slavin2009} as the canonical variables $(p,\phi)$ defined in Eq. (\ref{apppphi}), we get the positive/negative damping factors:
\setlength\abovedisplayskip{6pt}
\setlength\belowdisplayskip{6pt}
\begin{eqnarray}
S(p)&=&\frac{2p}{\alpha\left(\frac{\partial H_{O}}{\partial p}\right)}=\frac{2p}{\omega_{0}+kp},\nonumber\\
\beta(p,\mu)&=&2a_{J}(1-b_{J}p)p,\nonumber
\label{}
\end{eqnarray}
where using $p=(1/k)(\dot{\phi}-\omega_{0})$ $S(\omega_{o}^{-1}(\dot{\phi}))$ and $\beta(\omega_{o}^{-1}(\dot{\phi}),\mu)$ are easily obtained, and thereby we get the force $F$ in Eq. (\ref{nonconserNewtonc}). Then, requiring $F(\dot{\phi}_{T})=0$ as well as $F^{(1)}(\dot{\phi}_{T})<0$, one can easily get the terminal velocity $\dot{\phi}_{T}=\omega_{0}+k((a_{J}-\alpha)/(a_{J}b_{J}))\approx\omega_{0}$ and its stability criterion: $k>(a_{J}/\alpha)[k-\omega_{0}b_{J}-2k(1-\alpha/b_{J})]$, meaning that $a_{J}\in(\alpha,\alpha/(1-b_{J}))$, $b_{J}>0$, and $\omega_{0}>-k$. If we make a substitution on velocity like $\dot{\phi}'=\dot{\phi}-\omega_{0}\pm|\delta|$, where $|\delta|\ll1$ and the sign $\pm$ indicates the blue/red frequency shift cases, respectively, then $p\approx(1/k)\dot{\phi}'$ and thereby $F(\dot{\phi}'=0)=2a_{J}$, which means that quasi-linear oscillators also have a dc negative damping force and therefore they can be regarded as a sort of TVM particle either, as shown in Fig. \ref{termivelomotion}(e). Moreover, as Fig. \ref{termivelomotion}(f) shows, due to an extremely huge $m_{\mathrm{eff}}$ of the van der pol oscillator with $k=10^{-5}$, compared to another with a strong coefficient $k=1$, there seems to be no transient evolution in such an oscillator. But, actually, if we zoom in the vertical axis, the transient evolution for the extremely quasi-linear case is still observable like a typical TVM particle, as displayed by the panel of Fig. \ref{termivelomotion}(e). Thus, it is obvious that its generated frequency and oscillating amplitude can be tuned through varying $L$ (or $C$) and current (negative damping), respectively, which is very different from the oscillators with a considerable non-linear frequency shift like STNOs whose negative damping like STT can induce the variations of both of them in the meantime. Additionally, just as pointed out in Appendix \ref{appa:4}, such a huge mass will have the TVM particles with weak interactions simplified to the Adler ones.

\subsection{\label{C}TVM Model for Analyzing a Coupled pair of PERP-STNOs}
The TVM model for an asymmetric pair of PERP-STNOs coupled by a dipolar interaction has the form of Eq. (\ref{nonconserNewtonb}), which is derived form the Landau-Lifshitz-Gilbert-Slonczewski(LLGS) equation with the STT effect (Eq. (\ref{LLGinLab})) and the related material parameters of a PERP-STNO has been introduced in Appendix \ref{appa:5}. Also, the positive/negative damping forces and the interactive ones expressed in configuration space has been written in Appendix \ref{appa:5}.
 Here, we assume this pair of STNOs have different sets of spin-polarization efficiencies, i.e. $(\Lambda_{1},P_{1})=(2,0.38)$ and $(\Lambda_{2},P_{2})=(1.8,0.44)$, and both of them have the same demagnetization factor $k=1$ and Gilbert damping constant $\alpha_{1}=\alpha_{2}=0.02$. In order to analyze the excitation as well as phase-locking of them, we have to make a change of variables $\phi_{+}\equiv\phi_{1}+\phi_{2}$ and $\phi_{-}\equiv\phi_{1}-\phi_{2}$, where Eq. (\ref{TVM})  becomes
\setlength\abovedisplayskip{6pt}
\setlength\belowdisplayskip{6pt}
\begin{eqnarray}
\ddot{\phi}_{+}
&=&-\left(\frac{\alpha_{\mathrm{eff}+}}{2}\right)\dot{\phi}_{+}+\beta'_{+}-g_{0+}\sin\phi_{+},\nonumber\\
\ddot{\phi}_{-}&=&-\left(\frac{\alpha_{\mathrm{eff}+}}{2}\right)\dot{\phi}_{-}+\beta'_{-}-g_{0-}\sin\phi_{-},
\label{TVMphaslock}
\end{eqnarray}
where
\setlength\abovedisplayskip{6pt}
\setlength\belowdisplayskip{6pt}
\begin{eqnarray}
\alpha_{\mathrm{eff}\pm}&\equiv&\alpha_{1}\left(1-\dot{\phi}_{1}^{2}\right)\pm\alpha_{2}\left(1-\dot{\phi}_{2}^{2}\right),
\nonumber\\
\beta'_{\pm}&\equiv&\beta_{\pm}-\left(\frac{\alpha_{\mathrm{eff}-}}{2}\right)\dot{\phi}_{\mp},\nonumber\\
\beta_{\pm}&\equiv&(1-\dot{\phi}_{1}^{2})\frac{a_{J10}P_{1}\Lambda_{1}^{2}}{(\Lambda_{1}^{2}+1)
+(\Lambda_{1}^{2}-1)(-\dot{\phi}_{1})}\nonumber\\
&&\pm(1-\dot{\phi}_{2}^{2})\frac{a_{J20}P_{2}\Lambda_{2}^{2}}{(\Lambda_{2}^{2}+1)
+(\Lambda_{2}^{2}-1)(-\dot{\phi}_{2})},\nonumber\\
g_{0+}&\equiv&3A_{\mathrm{disc}}(d_{\mathrm{ee}})\sqrt{\left(1-\dot{\phi}_{1}^{2}\right)
\left(1-\dot{\phi}_{2}^{2}\right)},\nonumber\\
g_{0-}&\equiv&A_{\mathrm{disc}}(d_{\mathrm{ee}})\sqrt{\left(1-\dot{\phi}_{1}^{2}\right)
\left(1-\dot{\phi}_{2}^{2}\right)},\nonumber
\nonumber
\label{}
\end{eqnarray}
where since the terms $A\sum_{l=1}^{2}(\partial q_{i}/\partial\phi_{l})\dot{\phi}_{l}$ given in Appendix \ref{appa:5} are related to $\dot{\phi}_{\pm}$, which will have no contributions to the critical excitations of PL and AS states, thus they have been dropped out here. Note that $\dot{\phi}_{1,2}$ have to be replaced by $\dot{\phi}_{1(2)}=(1/2)(\dot{\phi}_{+}\pm\dot{\phi}_{-})$.

\subsubsection{\label{C1}Threshold Currents for PL States}
Just as pointed out in Ref. \cite{Chen2021}, the first and second equations in Eq.(\ref{TVMphaslock}) govern the excitations of the phase-locked (PL) states for the parallel and serial connected PERP-STNO pair, respectively, whose schematics of the structures can be seen in Ref. \cite{Chen2021}. Then, near the threshold currents driving the PL states, it is reasonably to assume that $\dot{\phi}_{1}\approx\pm\dot{\phi}_{2}$, i.e. $\dot{\phi}_{\mp}\approx0$ for the parallel and serial cases, respectively. Thereby, the equations for trigging the synchronized precessions of the STNO pairs, i.e. PL states, become
\setlength\abovedisplayskip{6pt}
\setlength\belowdisplayskip{6pt}
\begin{eqnarray}
\ddot{\phi}_{\pm}
&\approx&-\left(\frac{\alpha_{\mathrm{eff}+}}{2}\right)\dot{\phi}_{\pm}+\beta'_{\pm}-g_{0\pm}\sin\phi_{\pm}.
\label{TVMforPL}
\end{eqnarray}
where
\setlength\abovedisplayskip{6pt}
\setlength\belowdisplayskip{6pt}
\begin{eqnarray}
\alpha_{\mathrm{eff}+}&=&2\alpha\left[1-\left(\frac{\dot{\phi}_{+}}{2}
\right)^{2}\right],\hspace{0.5cm}(\textrm{parallel case})
\nonumber\\
&=&2\alpha\left[1-\left(\frac{\dot{\phi}_{-}}{2}
\right)^{2}\right],\hspace{0.5cm}(\textrm{serial case})
\nonumber\\
\beta'_{+}&=&a_{J0}\left[1-\left(\frac{\dot{\phi}_{+}}{2}
\right)^{2}\right]\Bigg[\frac{P_{1}\Lambda_{1}^{2}}{(\Lambda_{1}^{2}+1)+(\Lambda_{1}^{2}-1)\left(\frac{-\dot{\phi}_{+}}{2}
\right)}\nonumber\\
&&+\frac{P_{2}\Lambda_{2}^{2}}{(\Lambda_{2}^{2}+1)+(\Lambda_{2}^{2}-1)\left(\frac{-\dot{\phi}_{+}}{2}
\right)}\Bigg],\hspace{0.2cm}(\textrm{parallel case})\nonumber\\
\beta'_{-}&=&a_{J0}\left[1-\left(\frac{\dot{\phi}_{-}}{2}
\right)^{2}\right]\Bigg[\frac{P_{1}\Lambda_{1}^{2}}{(\Lambda_{1}^{2}+1)+(\Lambda_{1}^{2}-1)\left(\frac{-\dot{\phi}_{-}}{2}
\right)}\nonumber\\
&&+\frac{P_{2}\Lambda_{2}^{2}}{(\Lambda_{2}^{2}+1)+(\Lambda_{2}^{2}-1)\left(\frac{\dot{\phi}_{-}}{2}
\right)}\Bigg],\hspace{0.5cm}(\textrm{serial case})\nonumber\\
g_{0+}&=&3A_{\mathrm{disc}}(d_{\mathrm{ee}})\left[1-\left(\frac{\dot{\phi}_{+}}{2}\right)^{2}\right],\hspace{0.5cm}(\textrm{parallel case})\nonumber\\
\textrm{and}&&\nonumber\\
g_{0-}&=&A_{\mathrm{disc}}(d_{\mathrm{ee}})\left[1-\left(\frac{\dot{\phi}_{-}}{2}\right)^{2}\right].\hspace{0.5cm}(\textrm{serial case})\nonumber
\label{}
\end{eqnarray}
where, we have assumed that $a_{J10}=a_{J20}=a_{J0}$ and $a_{J10}=-a_{J20}=a_{J0}$ for the parallel and serial cases, respectively, which correspond to $I_{1}=I_{2}=I$ and $I_{1}=-I_{2}=I$, respectively.

We know that for the \emph{local} oscillatory states defined by the \textit{quasi-energy conservation} law (see Appendix \ref{appa:2}) is
\setlength\abovedisplayskip{6pt}
\setlength\belowdisplayskip{6pt}
\begin{eqnarray}
E_{\pm}=\frac{1}{2}\dot{\phi}_{\pm}^{2}-g_{0\pm}(\dot{\phi}_{\pm})\cos\phi_{\pm}
\label{quasienergy}
\end{eqnarray}
with $E_{\pm}<g_{0\pm}(\dot{\phi}_{\pm}=0)$, where $g_{0\pm}(\dot{\phi}_{\pm}=0)$ are the maxima of the potential energy, and the impacts of the forces $\beta'_{\pm}$ on these states are \textit{conservative-like} (see Ref. \cite{Chen2021}), meaning that $\beta'_{\pm}$ can drive the PL states through eliminating these local oscillatory states. Thus, one can solve the threshold currents $|I_{\mathrm{c,p(s)}}|$, where the subscripts p and s denote the parallel and serial cases, respectively, requiring
\setlength\abovedisplayskip{6pt}
\setlength\belowdisplayskip{6pt}
\begin{eqnarray}
\bigg|\beta'_{\pm}\left(\dot{\phi}_{\pm}=0\right)\bigg|
&=&g_{0\pm}\left(\dot{\phi}_{\pm}=0\right),
\label{TVMforIcps}
\end{eqnarray}
where the signs $\pm$ indicate the parallel and serial cases, respectively.

For the \emph{global} oscillatory states defined by Eq. (\ref{quasienergy}), i.e. the unperturbed PL states with $E_{\pm}>g_{0\pm}(\dot{\phi}_{\pm}=0)$, the effects of the forces $\beta_{\pm}$ on them are \emph{non-conservative}, where there must exist smaller threshold currents $|I_{\mathrm{b,p(s)}}|$ than $|I_{\mathrm{c,p(s)}}|$ to excite the PL states. The reason  for this is that the global oscillatory sates have more sufficient kinetic energies than those of the local oscillatory states, avoiding the particle from being trapped by the potential well\cite{Chen2021}. This also implies that there must be coexistent states S/PL that appears within $|I_{\mathrm{b,p(s)}}|<|I|<|I_{\mathrm{c,p(s)}}|$, where S means the static state with $\dot{\phi}_{1,2}=0$. Note that, these threshold currents $|I_{\mathrm{b,p(s)}}|$ are about the current-driven PL states whose energies are slightly  above $g_{0\pm}(\dot{\phi}_{\pm}=0)$. Then, based on the non-conservative effect of $\beta_{\pm}$ on the PL states, calculating the energy balance for them $|I_{\mathrm{b,p(s)}}|$ can be obtained as follows:  Firstly, the velocities of the unperturbed trajectories of the PL states can be obtained from Eq. (\ref{quasienergy}):
\setlength\abovedisplayskip{6pt}
\setlength\belowdisplayskip{6pt}
\begin{eqnarray}
\dot{\phi}_{+}&=&\pm\Bigg\{\frac{E_{+}+3A_{\mathrm{disc}}(d_{\mathrm{ee}})\cos\phi_{+}}{\frac{1}{2}-\frac{3}{4}A_{\mathrm{disc}}
(d_{\mathrm{ee}})\cos\phi_{+}}\Bigg\}^{1/2} \hspace{0.1cm}(\textrm{parallel case}),\nonumber\\
\dot{\phi}_{-}&=&\pm\Bigg\{\frac{E_{-}+A_{\mathrm{disc}}(d_{\mathrm{ee}})\cos\phi_{-}}{\frac{1}{2}-\frac{1}{4}A_{\mathrm{disc}}
(d_{\mathrm{ee}})\cos\phi_{-}}\Bigg\}^{1/2} \hspace{0.1cm}(\textrm{serial case}).
\label{velociunper}
\end{eqnarray}
Secondly, using Eqs. (\ref{velociunper}) and (\ref{TVMforPL}) one can calculate the time-averaged energy balance equations for the unperturbed PL states (see also Appendix. \ref{appa:3}):
\setlength\abovedisplayskip{6pt}
\setlength\belowdisplayskip{6pt}
\begin{eqnarray}
\langle\dot{E}_{\pm}\rangle_{T}&=&\frac{1}{T(E_{\pm})}\oint_{C(E_{\pm})}\left(\frac{d\phi_{\pm}}{\dot{\phi}_{\pm}}\right)
\dot{E}_{\pm},\nonumber\\
&=&\frac{1}{T(E_{\pm})}\int_{0(2\pi)}^{2\pi(0)}d\phi_{\pm}\left[-\frac{\alpha_{\mathrm{eff}+}\left(\dot{\phi}_{\pm}\right)}{2}\right]
\dot{\phi}_{\pm}\nonumber\\
&&+\frac{1}{T(E_{\pm})}\int_{0(2\pi)}^{2\pi(0)}d\phi_{\pm}\beta'_{\pm}\left(\dot{\phi}_{\pm}\right),
\label{energbalandPL}
\end{eqnarray}
where the periods of the unperturbed PL states are $T(E_{\pm})=\oint_{C(E_{\pm})}(d\phi_{\pm}/\dot{\phi}_{\pm})$, and $\phi_{\pm}\in[0(2\pi),2\pi(0)]$ which appears in the upper and lower limits of the integral means that $\dot{\phi}_{\pm}$ is positive or negative, respectively.  Finally, requiring $\langle\dot{E}_{\pm}(E_{\pm}\rightarrow g_{0\pm}(\dot{\phi}_{\pm}=0))\rangle_{T}=0$, we can solve $|I_{\mathrm{b,p(s)}}|$, which is proportional to the damping constant $\alpha$.
When $\alpha$ increases to certain critical values, i.e. $\alpha_{\mathrm{c,p(s)}}$, making $I_{\mathrm{c,p(s)}}=I_{\mathrm{b,p(s)}}$, then S/PL states will disappear. Notably, once $g_{0\pm}$ becomes smaller with an increasing $d_{\mathrm{ee}}$, $\alpha$ is also likely to become lager than $\alpha_{\mathrm{c,p(s)}}$ to have the S/PL states disappear. By the way, Eq. (\ref{TVMforPL}) has the four ingredients of a TVM particle mentioned previously, meaning that the triggered PL states are all stable.

As a small conclusion, we know that when $|I|>|I_{\mathrm{c,p(s)}}|$ the STNOs must be driven into the PL state no matter what the initial states are, which is owning to the elimination of the local oscillatory states by the negative damping force. However, when $|I_{\mathrm{c,p(s)}}|<|I|<|I_{\mathrm{c,p(s)}}|$, we know that only at those initial states with sufficient potential or kinetic energies can the pair of oscillators be driven into the PL states; otherwise their final states can be only the S states, which is so called IC-dependent excitations.

\subsubsection{\label{C2}Threshold Currents for AS States}
 One can solve the critical currents for trigging asynchronized states (AS), where $\phi_{\mp}$ increase with time for the parallel and serial cases, respectively, using the similar technique as solving the threshold currents for PL states. If the STNOs are in the PL regime ($\dot{\phi}_{\mp}\approx0$), then $\sin\phi_{\pm}$
must be a fast time-oscillating term such that $\langle\sin\phi_{\pm}\rangle_{T}\approx0$ in the time order of
the locked angles $\phi_{\mp}$ for the parallel and serial cases, respectively. Thus, at the stable states, i.e. $\ddot{\phi}_{\pm}\approx0$, Eq. (\ref{TVMforPL}) reduces to
\setlength\abovedisplayskip{6pt}
\setlength\belowdisplayskip{6pt}
\begin{eqnarray}
\left(\frac{\alpha_{\mathrm{eff}+}}{2}\right)\dot{\phi}_{\pm}=\beta'_{\pm}(\dot{\phi}_{\pm}),
\label{eqforIcpsaa}
\end{eqnarray}
where
\setlength\abovedisplayskip{6pt}
\setlength\belowdisplayskip{6pt}
\begin{eqnarray}
\alpha_{\mathrm{eff}+}&=&2\alpha\left[1-\left(\frac{\dot{\phi}_{+}}{2}
\right)^{2}\right],\hspace{0.3cm}(\textrm{parallel case})
\nonumber\\
&=&2\alpha\left[1-\left(\frac{\dot{\phi}_{-}}{2}
\right)^{2}\right],\hspace{0.3cm}(\textrm{serial case})
\nonumber\\
\beta'_{+}&=&a_{J0}\left[1-\left(\frac{\dot{\phi}_{+}}{2}
\right)^{2}\right]\Bigg[\frac{P_{1}\Lambda_{1}^{2}}{(\Lambda_{1}^{2}+1)+(\Lambda_{1}^{2}-1)\left(\frac{-\dot{\phi}_{+}}{2}
\right)}\nonumber\\
&&+\frac{P_{2}\Lambda_{2}^{2}}{(\Lambda_{2}^{2}+1)+(\Lambda_{2}^{2}-1)\left(\frac{-\dot{\phi}_{+}}{2}
\right)}\Bigg],\hspace{0.3cm}(\textrm{parallel case})\nonumber\\
\textrm{and}&&\nonumber\\
\beta'_{-}&=&a_{J0}\left[1-\left(\frac{\dot{\phi}_{-}}{2}
\right)^{2}\right]\Bigg[\frac{P_{1}\Lambda_{1}^{2}}{(\Lambda_{1}^{2}+1)+(\Lambda_{1}^{2}-1)\left(\frac{-\dot{\phi}_{-}}{2}
\right)}\nonumber\\
&&+\frac{P_{2}\Lambda_{2}^{2}}{(\Lambda_{2}^{2}+1)+(\Lambda_{2}^{2}-1)\left(\frac{\dot{\phi}_{-}}{2}
\right)}\Bigg].\hspace{0.3cm}(\textrm{parallel case})\nonumber
\label{}
\end{eqnarray}
Moreover, due to $\dot{\phi}_{\mp}\approx0$, one obtains the equations governing the phase-locked angles $\phi_{\mp}$ for the parallel and serial cases, respectively, from Eq. (\ref{TVMphaslock}):
\setlength\abovedisplayskip{6pt}
\setlength\belowdisplayskip{6pt}
\begin{eqnarray}
\ddot\phi_{\mp}=\beta'_{\mp}-g_{0\mp}\sin\phi_{\mp},
\label{eqforIcpsb}
\end{eqnarray}
where
\setlength\abovedisplayskip{6pt}
\setlength\belowdisplayskip{6pt}
\begin{eqnarray}
\alpha_{\mathrm{eff}-}&=&0,\hspace{0.3cm}(\textrm{parallel case})
\nonumber\\
&=&0,\hspace{0.3cm}(\textrm{serial case})
\nonumber\\
\beta'_{-}&=&a_{J0}\left[1-\left(\frac{\dot{\phi}_{+}}{2}
\right)^{2}\right]\Bigg[\frac{P_{1}\Lambda_{1}^{2}}{(\Lambda_{1}^{2}+1)+(\Lambda_{1}^{2}-1)\left(\frac{-\dot{\phi}_{+}}{2}
\right)}\nonumber\\
&&-\frac{P_{2}\Lambda_{2}^{2}}{(\Lambda_{2}^{2}+1)+(\Lambda_{2}^{2}-1)\left(\frac{-\dot{\phi}_{+}}{2}
\right)}\Bigg],\hspace{0.3cm}(\textrm{parallel case})\nonumber\\
\beta'_{+}&=&a_{J0}\left[1-\left(\frac{\dot{\phi}_{-}}{2}
\right)^{2}\right]\Bigg[\frac{P_{1}\Lambda_{1}^{2}}{(\Lambda_{1}^{2}+1)+(\Lambda_{1}^{2}-1)\left(\frac{-\dot{\phi}_{-}}{2}
\right)}\nonumber\\
&&-\frac{P_{2}\Lambda_{2}^{2}}{(\Lambda_{2}^{2}+1)+(\Lambda_{2}^{2}-1)\left(\frac{\dot{\phi}_{-}}{2}
\right)}\Bigg],\hspace{0.3cm}(\textrm{serial case})\nonumber\\
g_{0-}&=&A_{\mathrm{disc}}(d_{\mathrm{ee}})\left[1-\left(\frac{\dot{\phi}_{+}}{2}\right)^{2}\right],\hspace{0.3cm}(\textrm{parallel case})\nonumber\\
\textrm{and}&&\nonumber\\
g_{0+}&=&3A_{\mathrm{disc}}(d_{\mathrm{ee}})\left[1-\left(\frac{\dot{\phi}_{-}}{2}\right)^{2}\right].\hspace{0.3cm}(\textrm{serial case})\nonumber
\label{}
\end{eqnarray}

Similarly, the energy conservation that the \textit{local} oscillatory states defined by Eq. (\ref{eqforIcpsb}) satisfy is:
\setlength\abovedisplayskip{6pt}
\setlength\belowdisplayskip{6pt}
\begin{eqnarray}
E_{\mp}=\frac{1}{2}\dot{\phi}_{\mp}^{2}-g_{0\mp}(\dot{\phi}_{\pm})\cos\phi_{\mp}
\label{quasienergya}
\end{eqnarray}
with $E_{\mp}<g_{0\mp}(\dot{\phi}_{\pm})$, where $g_{0\mp}(\dot{\phi}_{\pm})$ are the maxima of the potential energies. The effects of the forces $\beta'_{\mp}$ on them are also conservative-like(see Ref. \cite{Chen2021}), indicating that the AS states can be excited through the removal of these local oscillating states by increasing current. Thus, the critical currents will satisfy:
\setlength\abovedisplayskip{6pt}
\setlength\belowdisplayskip{6pt}
\begin{eqnarray}
\bigg|\beta'_{\mp}\left(\dot{\phi}_{\pm}\right)\bigg|=g_{0\mp}\left(\dot{\phi}_{\pm}\right).
\label{eqforIcpsbb}
\end{eqnarray}

The procedure of obtaining the critical currents $|I'_{\mathrm{c,p(s)}}|$ for driving AS states is given as follows: Firstly, $a_{J0}$ as a function of $\dot{\phi}_{\pm}$ can be obtained using Eq. (\ref{eqforIcpsaa}), where $\dot{\phi}_{\pm}\in[-2,2]$ according to Eq. (\ref{dotphiasafuncp}); secondly, substituting $a_{J0}(\dot{\phi}_{\pm})$ into Eq. (\ref{eqforIcpsbb}) one solves the phase-locked angular velocities $\dot{\phi}_{\pm\mathrm{PL}}$; finally, substituting $\dot{\phi}_{\pm\mathrm{PL}}$ back into Eq. (\ref{eqforIcpsaa}) the critical currents can be easily obtained.

For the \textit{global} oscillatory states defined by Eq. (\ref{quasienergya}), i.e. the unperturbed AS states, the equations for triggering the AS states from the PL states are a little different from Eqs. (\ref{eqforIcpsaa}) and (\ref{eqforIcpsb}), where the constrains $\dot{\phi}_{\mp}=0$ at each moment during the whole excitation should be replaced by $\langle\dot{\phi}_{\mp}\rangle_{T}\rightarrow0$ when the critical phenomenon occurs between PL and PL/AS states. Thus, for this situation Eqs. (\ref{eqforIcpsaa}) and (\ref{eqforIcpsb}) will be replaced with
\setlength\abovedisplayskip{6pt}
\setlength\belowdisplayskip{6pt}
\begin{eqnarray}
\left(\frac{\alpha_{\mathrm{eff}+}}{2}\right)\dot{\phi}_{\pm}=\beta'_{\pm}(\dot{\phi}_{\pm}),
\label{eqforIcpsaa1}
\end{eqnarray}
where
\setlength\abovedisplayskip{6pt}
\setlength\belowdisplayskip{6pt}
\begin{eqnarray}
\alpha_{\mathrm{eff}+}&=&2\alpha\left[1-\left(\frac{\dot{\phi}_{+}}{2}
\right)^{2}\right],\hspace{0.3cm}(\textrm{parallel case})
\nonumber\\
&=&2\alpha\left[1-\left(\frac{\dot{\phi}_{-}}{2}
\right)^{2}\right],\hspace{0.3cm}(\textrm{serial case})
\nonumber\\
\beta'_{+}&\approx&a_{J0}\left[1-\left(\frac{\dot{\phi}_{+}}{2}
\right)^{2}\right]\Bigg[\frac{P_{1}\Lambda_{1}^{2}}{(\Lambda_{1}^{2}+1)+(\Lambda_{1}^{2}-1)\left(\frac{-\dot{\phi}_{+}}{2}
\right)}\nonumber\\
&&+\frac{P_{2}\Lambda_{2}^{2}}{(\Lambda_{2}^{2}+1)+(\Lambda_{2}^{2}-1)\left(\frac{-\dot{\phi}_{+}}{2}
\right)}\Bigg],\hspace{0.3cm}(\textrm{parallel case})\nonumber\\
\textrm{and}&&\nonumber\\
\beta'_{-}&\approx&a_{J0}\left[1-\left(\frac{\dot{\phi}_{-}}{2}
\right)^{2}\right]\Bigg[\frac{P_{1}\Lambda_{1}^{2}}{(\Lambda_{1}^{2}+1)+(\Lambda_{1}^{2}-1)\left(\frac{-\dot{\phi}_{-}}{2}
\right)}\nonumber\\
&&+\frac{P_{2}\Lambda_{2}^{2}}{(\Lambda_{2}^{2}+1)+(\Lambda_{2}^{2}-1)\left(\frac{\dot{\phi}_{-}}{2}
\right)}\Bigg],\hspace{0.3cm}(\textrm{serial case})\nonumber
\label{}
\end{eqnarray}

and
\setlength\abovedisplayskip{6pt}
\setlength\belowdisplayskip{6pt}
\begin{eqnarray}
\ddot\phi_{\mp}=-\left(\frac{\alpha_{\mathrm{eff}+}}{2}\right)\dot{\phi}_{\mp}+\beta'_{\mp}-g_{0\mp}\sin\phi_{\mp},
\label{eqforIcpsb1}
\end{eqnarray}
where
\setlength\abovedisplayskip{6pt}
\setlength\belowdisplayskip{6pt}
\begin{eqnarray}
\dot{\phi}_{1,(2)}&=&\frac{1}{2}\left(\dot{\phi}_{+}\pm\dot{\phi}_{-}\right),\nonumber\\
\alpha_{\mathrm{eff}\pm}&\equiv&\alpha\left[(1-\dot{\phi}_{1}^{2})\pm(1-\dot{\phi}_{2}^{2})\right],\nonumber\\
\beta'_{-}&=&\beta_{-}-\left(\frac{\alpha_{\mathrm{eff}-}}{2}\right)\dot{\phi}_{+},\hspace{0.3cm}(\textrm{parallel case})\nonumber\\
\beta'_{+}&=&\beta_{+}-\left(\frac{\alpha_{\mathrm{eff}-}}{2}\right)\dot{\phi}_{-},\hspace{0.3cm}(\textrm{serial case})\nonumber\\
\beta_{\mp}&\equiv&a_{J0}\Bigg[(1-\dot{\phi}_{1}^{2})\frac{P_{1}\Lambda_{1}^{2}}{(\Lambda_{1}^{2}+1)
+(\Lambda_{1}^{2}-1)(-\dot{\phi}_{1})}\nonumber\\
&&-(1-\dot{\phi}_{2}^{2})\frac{P_{2}\Lambda_{2}^{2}}{(\Lambda_{2}^{2}+1)
+(\Lambda_{2}^{2}-1)(-\dot{\phi}_{2})}\Bigg],\nonumber\\
g_{0-}&\approx&A_{\mathrm{disc}}(d_{\mathrm{ee}})\left[1-\left(\frac{\dot{\phi}_{+}}{2}\right)^{2}\right],\hspace{0.3cm}(\textrm{parallel case})\nonumber\\
\textrm{and}&&\nonumber\\
g_{0+}&\approx&3A_{\mathrm{disc}}(d_{\mathrm{ee}})\left[1-\left(\frac{\dot{\phi}_{-}}{2}\right)^{2}\right],\hspace{0.3cm}(\textrm{serial case})\nonumber
\label{}
\end{eqnarray}
respectively.

Following solving the threshold currents $|I_{\mathrm{b,p(s)}}|$ for the PL states in section \ref{C1}, similarly, the critical currents $|I'_{\mathrm{b,p(s)}}|$ for the driven AS states can be obtained as follows: Firstly, according to Eq. (\ref{quasienergya}), the velocities of the unperturbed AS states with the lower bound energies $E_{\mp}=g_{0\mp}$ are
\setlength\abovedisplayskip{6pt}
\setlength\belowdisplayskip{6pt}
\begin{eqnarray}
\dot{\phi}_{-}&=&\pm\sqrt{2E_{-}+2g_{0-}(\dot{\phi}_{+})\cos\phi_{-}}\hspace{0.3cm}(\textrm{parallel case}),\nonumber\\
\dot{\phi}_{+}&=&\pm\sqrt{2E_{+}+2g_{0+}(\dot{\phi}_{-})\cos\phi_{+}}\hspace{0.3cm}(\textrm{serial case}).
\label{velociunper1}
\end{eqnarray}

Secondly, using Eqs. (\ref{eqforIcpsb1}) and (\ref{velociunper1}) one can calculate the time-averaged energy balance equations for the unperturbed AS states:
\setlength\abovedisplayskip{6pt}
\setlength\belowdisplayskip{6pt}
\begin{eqnarray}
\langle\dot{E}_{\mp}\rangle_{T}&=&\frac{1}{T(E_{\mp})}\oint_{C(E_{\mp})}\left(\frac{d\phi_{\mp}}{\dot{\phi}_{\mp}}\right)
\dot{E}_{\mp},\nonumber\\
&=&\frac{1}{T(E_{\mp})}\int_{0(2\pi)}^{2\pi(0)}d\phi_{\mp}\left(-\frac{\alpha_{\mathrm{eff}+}}{2}\right)
\dot{\phi}_{\mp}\nonumber\\
&&+\frac{1}{T(E_{\mp})}\int_{0(2\pi)}^{2\pi(0)}d\phi_{\mp}\beta_{\mp}\nonumber\\
&&+\frac{1}{T(E_{\mp})}\int_{0(2\pi)}^{2\pi(0)}d\phi_{\mp}\left(-\frac{\alpha_{\mathrm{eff}-}}{2}\right)
\dot{\phi}_{\pm},
\label{energbalandPL1}
\end{eqnarray}
where the periods of the unperturbed AS states are $T(E_{\mp})=\oint_{C(E_{\mp})}(d\phi_{\mp}/\dot{\phi}_{\mp})$, and $\phi_{\mp}\in[0(2\pi),2\pi(0)]$ appearing in the upper and lower limits of the integral means that $\dot{\phi}_{\mp}$ is positive or negative, respectively. Then,
requiring $\langle\dot{E}_{\mp}(E_{\pm}\rightarrow g_{0\mp}(\dot{\phi}_{\pm}))\rangle_{T}=0$, we have
\setlength\abovedisplayskip{6pt}
\setlength\belowdisplayskip{6pt}
\begin{eqnarray}
a_{J0}=\frac{q_{\mp}(\dot{\phi}_{\pm})}{\zeta_{\mp}(\dot{\phi}_{\pm})},
\label{I1}
\end{eqnarray}
where
\setlength\abovedisplayskip{6pt}
\setlength\belowdisplayskip{6pt}
\begin{eqnarray}
q_{\mp}(\dot{\phi}_{\pm})&\equiv&-\int_{0(2\pi)}^{2\pi(0)}d\phi_{\mp}\left[\left(-\frac{\alpha_{\mathrm{eff}+}}{2}\right)
\dot{\phi}_{\mp}+\left(-\frac{\alpha_{\mathrm{eff}-}}{2}\right)\dot{\phi}_{\pm}\right],\nonumber\\
\zeta_{\mp}(\dot{\phi}_{\pm})&=&\left(\frac{1}{a_{J0}}\right)\int_{0(2\pi)}^{2\pi(0)}d\phi_{\mp}\beta_{\mp},\nonumber
\end{eqnarray}
 and $\dot{\phi}_{\pm}\in[-2,2]$.

Thirdly, from Eqs. (\ref{eqforIcpsaa1}) and (\ref{I1}), one gets $I_{\mp}=I_{\mp}(\dot{\phi}_{\pm})$ and $I'_{\mp}=I'_{\mp}(\dot{\phi}_{\pm})$, respectively. Then, requiring $I_{\mp}=I'_{\mp}$, the velocity $\dot{\phi'}_{\pm\mathrm{PL}}$ can be solved. Finally, substituting $\dot{\phi'}_{\pm\mathrm{PL}}$ back into Eq. (\ref{eqforIcpsaa1}) one obtains the critical currents $|I'_{\mathrm{b,p(s)}}|$. Note, that, similar to the discussion about the presence of the S/PL regime, due to higher kinetic energy of the AS states, there must be coexistent states PL/AS appearing within $|I'_{\mathrm{b,p(s)}}|<|I|<|I'_{\mathrm{c,p(s)}}|$. Once $\alpha$ exceeds certain critical values, i.e. $\alpha'_{\mathrm{c,p(s)}}$, or if $g_{0\mp}$ becomes smaller with an increasing $d_{\mathrm{ee}}$ making $I'_{\mathrm{c,p(s)}}=I'_{\mathrm{b,p(s)}}$ then PL/AS states will disappear.

By the way, the dynamic behavior of the driven AS states are all stable according to Eq. (\ref{eqforIcpsb1}), which is also a typical TVM particle that has been introduced in Sec. \ref{B}.

\subsubsection{\label{C3}Phase-locked Frequencies and Phase Angles}
The phase-locked frequencies as a function of current $I$ can be obtained as follows. Using Eqs. (\ref{velociunper}) and (\ref{energbalandPL}), giving a value of $I$ from the range $|I_{\mathrm{b,p(s)}}|<|I|<|I'_{\mathrm{c,p(s)}}|$, and requiring $\langle\dot{E}_{\pm}(E_{\pm\mathrm{PL}})\rangle_{T}=0$, the solution $E_{\pm\mathrm{PL}}(I)$ used to designate the driven PL state can be obtained. Then, one have
\setlength\abovedisplayskip{6pt}
\setlength\belowdisplayskip{6pt}
\begin{eqnarray}
f_{\mathrm{PL}}(\mathrm{GHz})&=&\left(\frac{4\pi M_{s}\gamma}{2\pi}\right)\left[\frac{1}{2}\langle\dot{\phi}_{\pm\mathrm{PL}}\rangle_{T}\right],\nonumber\\
&=&\left(\frac{4\pi M_{s}\gamma}{2\pi}\right)\left(\frac{2\pi}{2}\right)\left|\int_{0(2\pi)}^{2\pi(0)}\frac{d\phi_{\pm}}{\dot{\phi}_{\pm}
(E_{\pm\mathrm{PL}})}\right|^{-1}.
\label{PLfreq}
\end{eqnarray}
Moreover, substituting $\langle\dot{\phi}_{\pm\mathrm{PL}}\rangle_{T}$ calculated above into Eq. (\ref{eqforIcpsb}) and requiring
$\beta'_{\mp}=g_{0\mp}\sin\phi_{\mp}$, the phase-locked angles as a function of current $\phi_{\mp\mathrm{PL}}(I)$ can be obtained to be
\setlength\abovedisplayskip{6pt}
\setlength\belowdisplayskip{6pt}
\begin{eqnarray}
\phi_{\mp\mathrm{PL}}=\sin^{-1}\left(\frac{\beta'_{\mp}}{g_{0\mp}}\right).
\label{PLangle}
\end{eqnarray}
By the way, the procedure of obtaining the phase-locked angles $\phi_{\mp\mathrm{PL}}$ here, where the angular velocities $\langle\dot{\phi}_{\pm\mathrm{PL}}\rangle_{T}$ have to be gotten first, basically corresponds to that of solving the phase-locked angle $\Phi_{0}$ in phase space for the injection-locking of a PERP-STNO, where the phase-locked momentum $p_{\mathrm{PL}}$ has to be gotten in advance, as has been discussed in Sec. \ref{A}.

\subsubsection{\label{C4}Asynchronized Frequencies}
In contrast to the PL regime, the asynchronized (AS) regime means that the pair of STNOs get rid of the phase-locking induced by a certain of coupling mechanism, that is, the operation of them would be certainly quite close to the free-running regime. Thus, the asynchronized frequencies $f_{1,2}$ for the pair of STNOs can be reasonably calculated through the linear expansion of the TVM model, i.e. Eq. (\ref{nonconserNewtonb}), about their respective free-running velocities $\dot{\phi}_{i0}$ as follows:
Firstly, Eq. (\ref{nonconserNewtonb}) can be rewritten to be
\setlength\abovedisplayskip{6pt}
\setlength\belowdisplayskip{6pt}
\begin{eqnarray}
\ddot{\phi}_{i}&=&F_{i}(\dot{\phi}_{j},\phi_{j}),\nonumber\\
&=&F_{i0}(\dot{\phi}_{i})+F_{iI}(\dot{\phi}_{j},\phi_{j}).
\label{LTVM1}
\end{eqnarray}
where the first and second terms appearing on the right-hand side of Eq. (\ref{LTVM1}) are the non-conservative forces supporting stable TVM and the conservative forces generating interactive forces, respectively. Thereby, in the absence of $F_{iI}$ , requiring $F_{i0}(\dot{\phi}_{i0})=0$ and $F_{i0}^{(1)}(\dot{\phi}_{i0})<0$ the stable TVM labeled by $\dot{\phi}_{i0}$ can be solved.

Furthermore, if a small $F_{iI}$ is present, the perturbation in $\dot{\phi}_{i}$  must be due to this anisotropic force depending on angles $\phi_{j}$. Thus, making a transformation of velocities $\dot{\phi}_{i}=\dot{\phi}_{i0}+\delta\dot{\phi}_{i}$ the linear expansion of Eq. (\ref{LTVM1}) about $\dot{\phi}_{i0}$ reads
\setlength\abovedisplayskip{6pt}
\setlength\belowdisplayskip{6pt}
\begin{eqnarray}
\delta\ddot{\phi}_{i}&\approx&F_{i0}^{(1)}(\dot{\phi}_{i0})\delta\dot{\phi}_{i}+F_{iI}(\dot{\phi}_{j0},\phi_{j}),
\label{LTVM2}
\end{eqnarray}
where $F_{iI}(\dot{\phi}_{j0},\phi_{j})\equiv-\partial U_{I}/\partial\phi_{i}$, $U_{I}=U_{I}(\dot{\phi}_{j0},\phi_{j})$, and the
first and higher order terms related to $\delta\dot{\phi}_{j}$ appearing in $U_{I}$ have been neglected.
According to the quasi-energy conservation, i.e. $E_{\mathrm{a}}(\dot{\phi}_{j0},\phi_{j\mathrm{a}},t_{\mathrm{a}})=E_{\mathrm{b}}(\dot{\phi}_{j0}+\delta\dot{\phi}_{j}
,\phi_{j\mathrm{b}},t_{\mathrm{b}})$, the order of magnitude of $\delta\dot{\phi}_{i}$ is estimated to be $|\delta\dot{\phi}_{i}|\sim\sqrt{2|\Delta U_{I}|}$, which is sufficiently smaller than $\dot{\phi}_{i0}$ to ensure Eq. (\ref{LTVM2}) is valid. And, finally, returning back to the original frame of reference, i.e. using $\delta\dot{\phi}_{i}=\dot{\phi}_{i}-\dot{\phi}_{i0}$, the linearized TVM model reads
\setlength\abovedisplayskip{6pt}
\setlength\belowdisplayskip{6pt}
\begin{eqnarray}
\ddot{\phi}_{i}&=&F_{i0}^{(1)}(\dot{\phi}_{i0})\dot{\phi}_{i}-F_{i0}^{(1)}(\dot{\phi}_{i0})\dot{\phi}_{i0}
+F_{iI}(\dot{\phi}_{j0},\phi_{j}),
\label{LTVM3}
\end{eqnarray}
where, notably, the first and second terms on the right-hand side are the positive and dc negative damping forces, respectively.

Moreover, Eq. (\ref{LTVM3}) for the coupled pair of PERP-STNOs in the AS regime becomes
\setlength\abovedisplayskip{6pt}
\setlength\belowdisplayskip{6pt}
\begin{eqnarray}
\ddot{\phi}_{1}&=&F_{10}^{(1)}(\dot{\phi}_{10})\dot{\phi}_{1}-F_{10}^{(1)}(\dot{\phi}_{10})\dot{\phi}_{10}
+F_{1I}(\dot{\phi}_{10},\dot{\phi}_{20},\phi_{1},\phi_{2}),\nonumber\\
\ddot{\phi}_{2}&=&F_{20}^{(1)}(\dot{\phi}_{20})\dot{\phi}_{2}-F_{20}^{(1)}(\dot{\phi}_{20})\dot{\phi}_{20}
+F_{2I}(\dot{\phi}_{10},\dot{\phi}_{20},\phi_{1},\phi_{2}),\nonumber\\
\label{LTVMforPERP-STNO}
\end{eqnarray}
where the stable free-running velocities $\dot{\phi}_{i0}$ and their stabilities $F_{i0}^{(1)}$ are given as follows:
\setlength\abovedisplayskip{6pt}
\setlength\belowdisplayskip{6pt}
\begin{eqnarray}
\dot{\phi}_{i0}=\frac{(\Lambda_{i}^{2}+1)-\sqrt{(\Lambda_{i}^{2}+1)^{2}-4(\Lambda_{i}^{2}-1)\left(a_{Ji0}P_{i}\Lambda_{i}^{2}
/\alpha_{i}\right)}}{2(\Lambda_{i}^{2}-1)},\nonumber
\label{}
\end{eqnarray}
and
\setlength\abovedisplayskip{6pt}
\begin{eqnarray}
F_{i0}^{(1)}(\dot{\phi}_{i0})&=&-\alpha_{i}(1-\dot{\phi}_{i0}^{2})\Bigg\{1-\left(\frac{a_{Ji0}P_{i}\Lambda_{i}^{2}}{\alpha_{i}}\right)\nonumber\\
&&\times\frac{(\Lambda_{i}^{2}-1)}{\left[(\Lambda_{i}^{2}+1)+(\Lambda_{i}^{2}-1)(-\dot{\phi}_{i0})\right]^{2}}\Bigg\},\nonumber
\label{}
\end{eqnarray}
and where the velocities $\dot{\phi}_{i}$ originally appearing in the interactive forces in Eq. (\ref{TVMphaslock}) have been replaced by their free-running values $\dot{\phi}_{i0}$ in $F_{iI}$. Then, introducing a change of variables $\phi_{\pm}=\phi_{1}\pm\phi_{2}$, Eq. (\ref{LTVMforPERP-STNO}) becomes
\setlength\abovedisplayskip{6pt}
\setlength\belowdisplayskip{6pt}
\begin{eqnarray}
\ddot{\phi}_{+}-F_{+}^{(1)}\dot{\phi}_{+}-F_{-}^{(1)}\dot{\phi}_{-}
&=&F_{\mathrm{dc}+}-g_{0+}\sin\phi_{+},\nonumber\\
\ddot{\phi}_{-}-F_{+}^{(1)}\dot{\phi}_{-}-F_{-}^{(1)}\dot{\phi}_{+}
&=&F_{\mathrm{dc}-}-g_{0-}\sin\phi_{-},
\label{LTVMforAS}
\end{eqnarray}
where $F_{\pm}^{(1)}\equiv(1/2)\left[F_{10}^{(1)}\pm F_{20}^{(1)}\right]$ , $F_{\mathrm{dc}\pm}\equiv\pm\left[F_{10}^{(1)}\dot{\phi}_{10}\pm F_{20}^{(1)}\dot{\phi}_{20}\right]$.

Having known that under the AS regime, i.e. $|I|>|I'_{\mathrm{b,p(s)}}|$, the first and second equations in Eq. (\ref{LTVMforAS}) govern the excitation of precession for the parallel (+) and serial (-) connections, respectively, one can reasonably assume that $|F_{\mathrm{dc}\pm}|\gg|g_{0\pm}|$ and $\langle\sin\phi_{\pm}\rangle_{T}\approx0$. Thus, for the AS states, i.e. $\ddot{\phi}_{\pm}\approx0$, these equations become
\setlength\abovedisplayskip{6pt}
\setlength\belowdisplayskip{6pt}
\begin{eqnarray}
\dot{\phi}_{\pm}=-\left[\frac{1}{F_{+}^{(1)}}\right]\left[F_{\mathrm{dc}\pm}+F_{-}^{(1)}\dot{\phi}_{\mp}\right].
\label{LTVMforprecess}
\end{eqnarray}
Furthermore, substituting Eq. (\ref{LTVMforprecess}) into the equations governing the phase-locking in Eq. (\ref{LTVMforAS}), we have
\setlength\abovedisplayskip{6pt}
\setlength\belowdisplayskip{6pt}
\begin{eqnarray}
\ddot{\phi}_{\mp}-F'^{(1)}_{+}\dot{\phi}_{\mp}=F'_{\mathrm{dc}\mp}-g_{0\mp}\sin\phi_{\mp},
\label{LTVMforAS1}
\end{eqnarray}
where $F'^{(1)}_{+}\equiv F^{(1)}_{+}-\left[\left(F_{-}^{(1)}\right)^{2}/F_{+}^{(1)}\right]$ and $F'_{\mathrm{dc}\mp}\equiv F_{\mathrm{dc}\mp}-\left[F_{-}^{(1)}/F_{+}^{(1)}\right]F_{\mathrm{dc}\pm}$.

Similar to obtaining the phase-locked frequency (see Sections \ref{C2} and \ref{C3}), the unperturbed AS states with energy $E_{\mp}>|g_{0\mp}|$ are
\setlength\abovedisplayskip{6pt}
\setlength\belowdisplayskip{6pt}
\begin{eqnarray}
\dot{\phi}_{-}&=&\pm\sqrt{2E_{\mp}+2g_{0-}(\dot{\phi}_{+})\cos\phi_{-}}\hspace{0.3cm}(\textrm{parallel case}),\nonumber\\
\dot{\phi}_{+}&=&\pm\sqrt{2E_{\mp}+2g_{0+}(\dot{\phi}_{-})\cos\phi_{+}}\hspace{0.3cm}(\textrm{serial case}).
\label{velociunperAS}
\end{eqnarray}
Then, due to the non-conservative force of $F'_{\mathrm{dc}\mp}$ for these states, the time-averaged energy balance equations for the unperturbed AS states are
\setlength\abovedisplayskip{6pt}
\setlength\belowdisplayskip{6pt}
\begin{eqnarray}
\langle\dot{E}_{\mp}\rangle_{T}&=&\frac{1}{T(E_{\mp\mathrm{AS}})}\oint_{C(E_{\mp\mathrm{AS}})}\left(\frac{d\phi_{\mp}}{\dot{\phi}
_{\mp}}\right)\dot{E}_{\mp},\nonumber\\
&=&\frac{1}{T(E_{\mp\mathrm{AS}})}\int_{0(2\pi)}^{2\pi(0)}d\phi_{\mp}\left[F'^{(1)}_{+}\right]
\dot{\phi}_{\mp}\nonumber\\
&&+\frac{1}{T(E_{\mp\mathrm{AS}})}\int_{0(2\pi)}^{2\pi(0)}d\phi_{\mp}F'_{\mathrm{dc}\mp}.\nonumber\\
\label{energbalandAS}
\end{eqnarray}
Requiring $\langle\dot{E}_{\mp}(E_{\mp\mathrm{AS}})\rangle_{T}=0$,  the solution $E_{\mp\mathrm{AS}}(I)$
used to label the driven AS state can be obtained. Thereby, from Eqs. (\ref{LTVMforprecess}) and (\ref{energbalandAS}) one gets the time-averaged velocities:
\setlength\abovedisplayskip{6pt}
\setlength\belowdisplayskip{6pt}
\begin{eqnarray}
\langle\dot{\phi}_{\pm}\rangle_{T}&=&-\left[\frac{1}{F_{+}^{(1)}}\right]\left[F_{\mathrm{dc}\pm}+F_{-}^{(1)}\langle\dot{\phi}
_{\mp}\rangle_{T}\right],\nonumber\\
\langle\dot{\phi}_{\mp}\rangle_{T}&=&\frac{2\pi}{\int_{0(2\pi)}^{(2\pi)0}\frac{d\phi_{\mp}}{\sqrt{2E_{\mp\mathrm{AS}}+2g_{0\mp}
\cos\phi_{\mp}}}}.\nonumber
\label{}
\end{eqnarray}
And, finally, we obtain the frequencies for the AS regime:
\setlength\abovedisplayskip{6pt}
\setlength\belowdisplayskip{6pt}
\begin{eqnarray}
f_{1}(\mathrm{GHz})&=&\frac{4\pi M_{s}\gamma}{2(2\pi)}\left|\langle\dot{\phi}_{+}\rangle_{T}+\langle\dot{\phi}_{-}\rangle_{T}\right|,\nonumber\\
f_{2}(\mathrm{GHz})&=&\frac{4\pi M_{s}\gamma}{2(2\pi)}\left|\langle\dot{\phi}_{+}\rangle_{T}-\langle\dot{\phi}_{-}\rangle_{T}\right|.
\label{}
\end{eqnarray}

By the way, the critical currents $|I'_{\mathrm{b,p(s)}}|$ between the PL and PL/AS states can also be well calculated from Eq. (\ref{LTVMforAS}) following the same way used in Sec. \ref{C2}.

\subsection{\label{D}Threshold and Critical Currents Confirmed by the TVM/Macrospin Simulations}
\begin{figure*}
\begin{center}
\includegraphics[width=16cm]{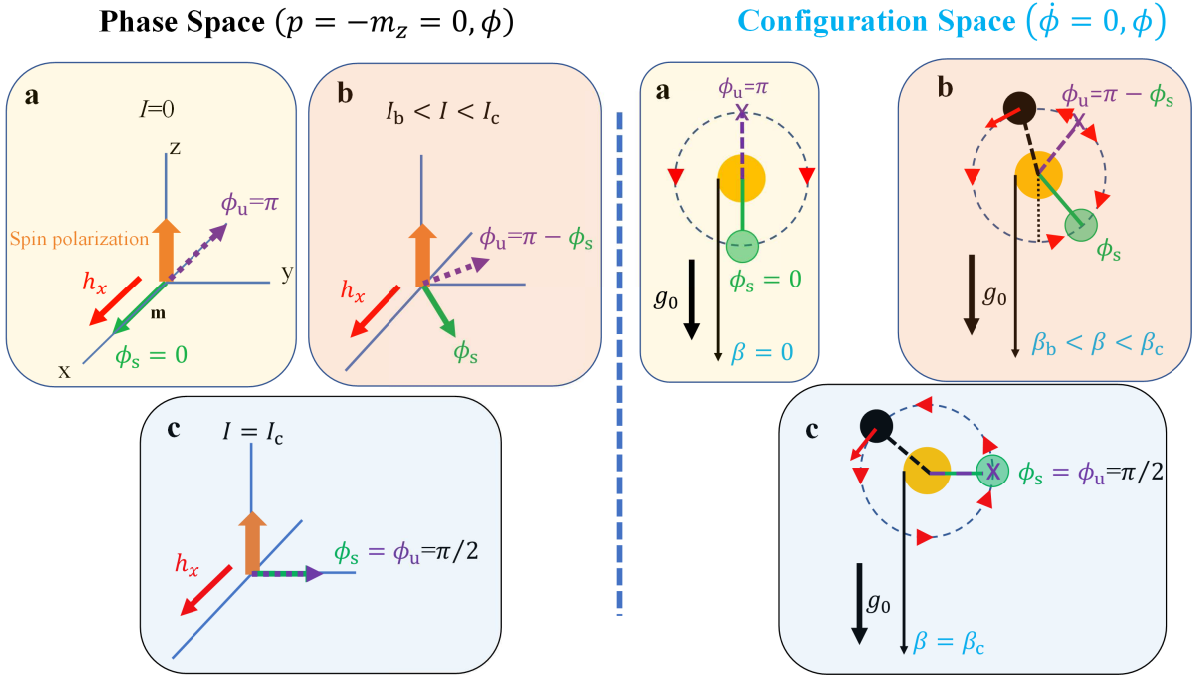}
\end{center}
\caption{(Color Online) Comparison of the threshold excitation of a PERP-STNO under a uniform magnetic field (left part) with that of a forced pendulum under a uniform gravitational field (right part). Figures \textbf{a}, \textbf{b}, and \textbf{c} indicate the cases of $I=0$ $(\beta=0)$, $I_{\mathrm{b}}<I<I_{\mathrm{c}}$ $(\beta_{\mathrm{b}}<\beta<\beta_{\mathrm{c}})$, and $I=I_{\mathrm{c}}$ ($\beta=\beta_{\mathrm{c}}$), respectively, where $I$ and $\beta$ are the injected current and driving force, respectively. In the left part, the orange, red, green and purple dot arrows indicate the spin polarization vector, the applied magnetic field $h_{x}$, the stable $\phi_{\mathrm{s}}$, and unstable $\phi_{\mathrm{u}}$ equilibrium points of the free layer moment, respectively. In the right part, the black and red arrows denote the uniform gravitational field $g_{0}$ and the acting forces on the pendulum represented by the black ball, respectively. Additionally, the green ball and purple 'X' mark the angles of stable and unstable equilibrium points, respectively.
}
\label{Pendulum}
\end{figure*}
We would like to stress that even though the analytical approach for the threshold and critical currents shown above looks a little complicated they can also be easily found out by performing the Macrospin or TVM simulations.
Just as pointed out previously, the dynamics of a phase-locked non-linear auto-oscillator can be well modeled by a forced pendulum \cite{Chen2021}, which is certainly a sort of a TVM particle in nature in the configuration space $(\dot{\phi},\phi)$. Thereby, analyzing a forced pendulum under a uniform gravitational force/field, which is the simplest form of an anisotropic force/field dependent on the phase angle $\phi$, can give appropriate initial conditions with zero initial velocities $\dot{\phi}_{0}=0$ to drive global oscillations, i.e. OP precessions in PERP-STNOs, as depicted in the right part of Fig. \ref{Pendulum}. As a comparison, in the left part of Fig. \ref{Pendulum}, the corresponding initial conditions in a PERP-STNO are also given in the phase space $(p=-m_{z},\phi)$. Moreover, the uniform gravitational force/field in configuration space can be replaced with a uniform magnetic field applied along the x axis in phase space.

In the absence of the driving force $\beta=0$ (or $I=0$), see Fig. \ref{Pendulum}\textbf{a}, there are a stable $\phi_{\mathrm{s}}$ and an unstable $\phi_{\mathrm{un}}$ equilibrium points that are parallel and anti-parallel to the gravitational force, respectively. Before $|\beta|$ ($|I|$) is increased up to $|\beta_{\mathrm{c}}|$ ( $|I_{\mathrm{c}}|$), as Figs. \ref{Pendulum}\textbf{b} and \textbf{c} display, both of the two points will gradually move close to $\pm\pi/2$, where $\phi_{\mathrm{un}}=\pi-\phi_{\mathrm{s}}$\cite{Chen2021}. If the pendulum is initially placed between them, then it will eventually evolve into the static state $\phi_{\mathrm{s}}$. Besides, if the pendulum is moved close enough to $\phi_{\mathrm{un}}$ from its left or right sides against the driving force $\beta$ for the cases of positive or negative $\beta$ (or $I$), respectively, the pendulum would probably gain enough kinetic energy from the driving force or the STT to trigger the global oscillations (or OP precessions in PERP-STNOs), which is termed a \textit{slingshot effect}. Therefore, using this effect, one can confirm whether the injected current can drive an OP state, helping us to find out $|\beta_{\mathrm{b}}|$ ($|I_{\mathrm{b}}|$) numerically in a more efficient way.

\begin{figure*}
\begin{center}
\includegraphics[width=13cm]{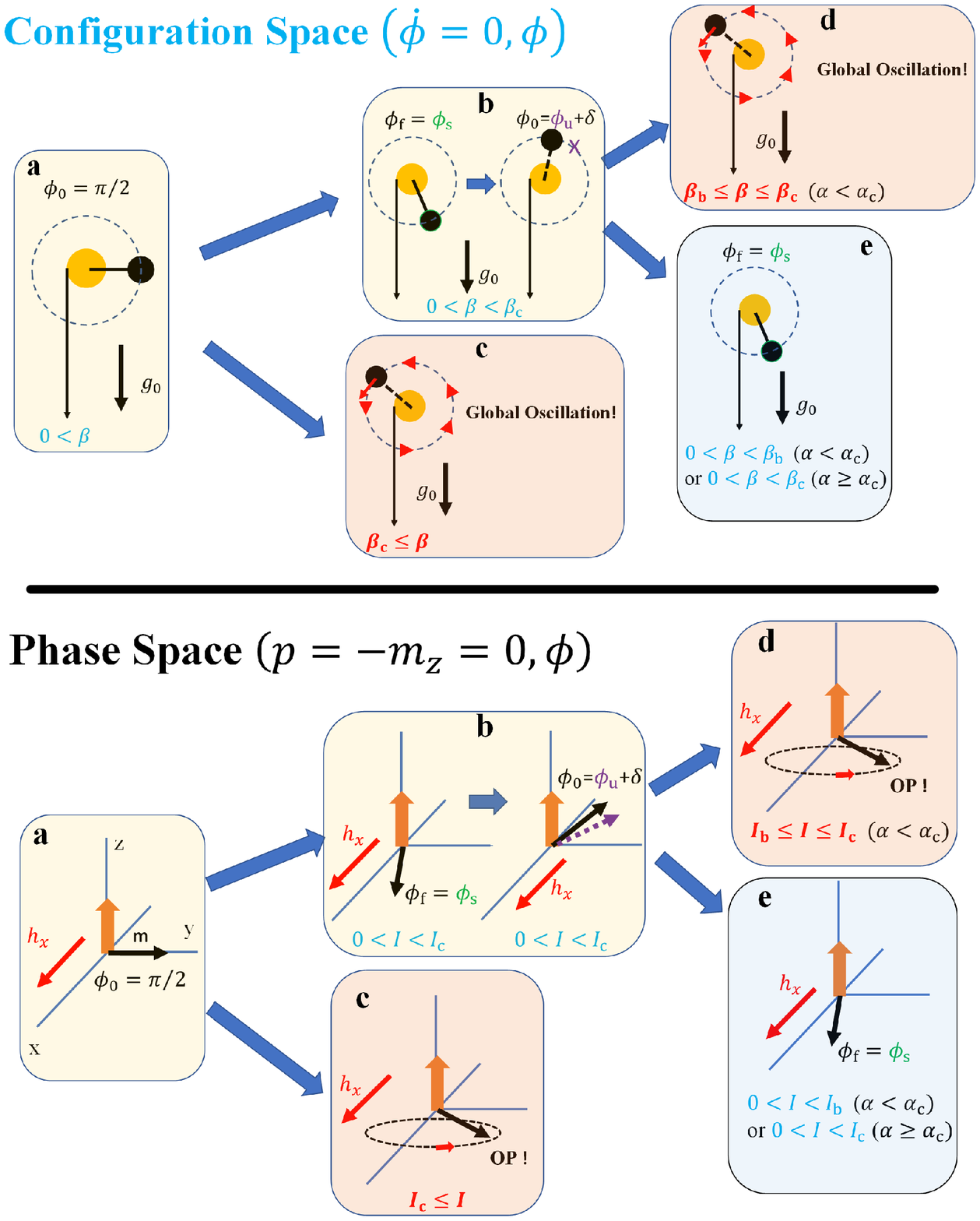}
\end{center}
\caption{(Color Online)  Standard operating procedure to search for the threshold currents $|I_\mathrm{b(c)}|$ (driving forces $|\beta_{\mathrm{b(c)}}|$) by performoing numerical simulations. Notably, the equivalent procedures demonstrated in the configuration and phase spaces, respectively, are both displayed here. Figure \textbf{a} indicates placing the initial phase angle at $\phi_{0}=\pi/2$ in the very beginning. Figure \textbf{b} shows how to check whether the current (or driving force) can drive an OP state (global oscillation) using the slingshot effect where $\delta$ is a very small amount of deviation after the stable state $\phi_{\mathrm{s}}$ has been gotten. Figure \textbf{d} shows the driven OP state by this effect. Figure \textbf{c} indicates the appearance of $|I_{\mathrm{c}}|$ ($|\beta_{\mathrm{c}}|$) when the stable and unstable states $\phi_{\mathrm{s(u)}}$ both disappear at $\phi=\pi/2$. Figure \textbf{e} stresses that whether the slingshot effect can trigger an OP state before the appearance of $|I_{\mathrm{c}}|$ manifests the cases of $\alpha<\alpha_{\mathrm{c}}$ or $\alpha\geq\alpha_{\mathrm{c}}$.}
\label{initialsetting}
\end{figure*}

Finally, if $|\beta|=|\beta_{\mathrm{c}}|$ (or $|I|=|I_{\mathrm{c}}|$), these two points $\phi_{\mathrm{s(un)}}$ will merge together into a single unstable equilibrium point $\phi=\pm\pi/2$, where the pendulum will eventually evolve into a global oscillation no matter what its initial states are, as presented by Fig. \ref{Pendulum} \textbf{c}. By the way, there exists an upper limit of the damping constant, i.e. $\alpha_{\mathrm{c}}=(2\pi/S')\sqrt{g_{0}/2}$, ensuring the existence of $\beta_{\mathrm{b}}$ (or $I_{\mathrm{b}}$), where $S'=\int_{0}^{2\pi}d\phi\sqrt{1+\cos\phi}\approx5.6569$ \cite{Chen2021}. So, placing the initial angle at $\pm\pi/2$ can help us not only  get $\phi_{\mathrm{s}}$ to find out $|I_{\mathrm{b}}|$ using the slingshot effect but also search for $|I_{\mathrm{c}}|$.

Having introduced the initial conditions for the threshold excitation of a single PERP-STNO (or a single pendulum), one can easily give a standard operating procedure to find out these threshold currents performing numerical simulations, as displayed in Fig. \ref{initialsetting}: Firstly, see Figs. \ref{initialsetting}  \textbf{a}, \textbf{b}, and \textbf{d}. The initial state of the moment (pendulum) has to be first set at  $(p_{0}=-m_{z0}=0,\phi_{0}=\pm\pi/2)$ ($(\dot{\phi}_{0}=0,\phi_{0}=\pm\pi/2)$) to make the system evolve into the stable state $(p=0,\phi_{\mathrm{s}})$ ($(\dot{\phi}=0,\phi_{\mathrm{s}})$), where the signs $\pm$ correspond to the cases of positive or negative $I$ ($\beta$), respectively.  Then, utilizing the slingshot effect and gradually tuning up the current (driving force) amplitude, the value of $|I_{\mathrm{b}}|$ ($|\beta_{\mathrm{b}}|$) could be found out till the excitation of OP states (global oscillations); Secondly, see Figs. \ref{initialsetting} \textbf{a} and \textbf{c}. The threshold current $|I_{\mathrm{c}}|$ ($|\beta_{\mathrm{c}}|$) can be easily gotten by setting the initial state at $(p_{0}=0,\phi_{0}=\pm\pi/2)$ ($(\dot{\phi}_{0}=0,\phi_{0}=\pm\pi/2)$) and then gradually increasing the current (driving force) amplitude till the trigging of OP states (global oscillations).

It should be noticed here that the above procedure about successful searching for $|I_{\mathrm{b}}|$ ($|\beta_{\mathrm{b}}|$), as mentioned previously, manifests $\alpha<\alpha_{\mathrm{c}}$. Thus, once $|I_{\mathrm{b}}|$ ($|\beta_{\mathrm{b}}|$) fails to be found out numerically using the slingshot effect before the appearance of $|I_{\mathrm{c}}|$ ($|\beta_{\mathrm{c}}|$), it means $\alpha\geq\alpha_{\mathrm{c}}$ instead, as shown in Figs. \ref{initialsetting}  \textbf{a}, \textbf{b}, and \textbf{e}.

\begin{figure*}
\begin{center}
\includegraphics[width=17.7cm]{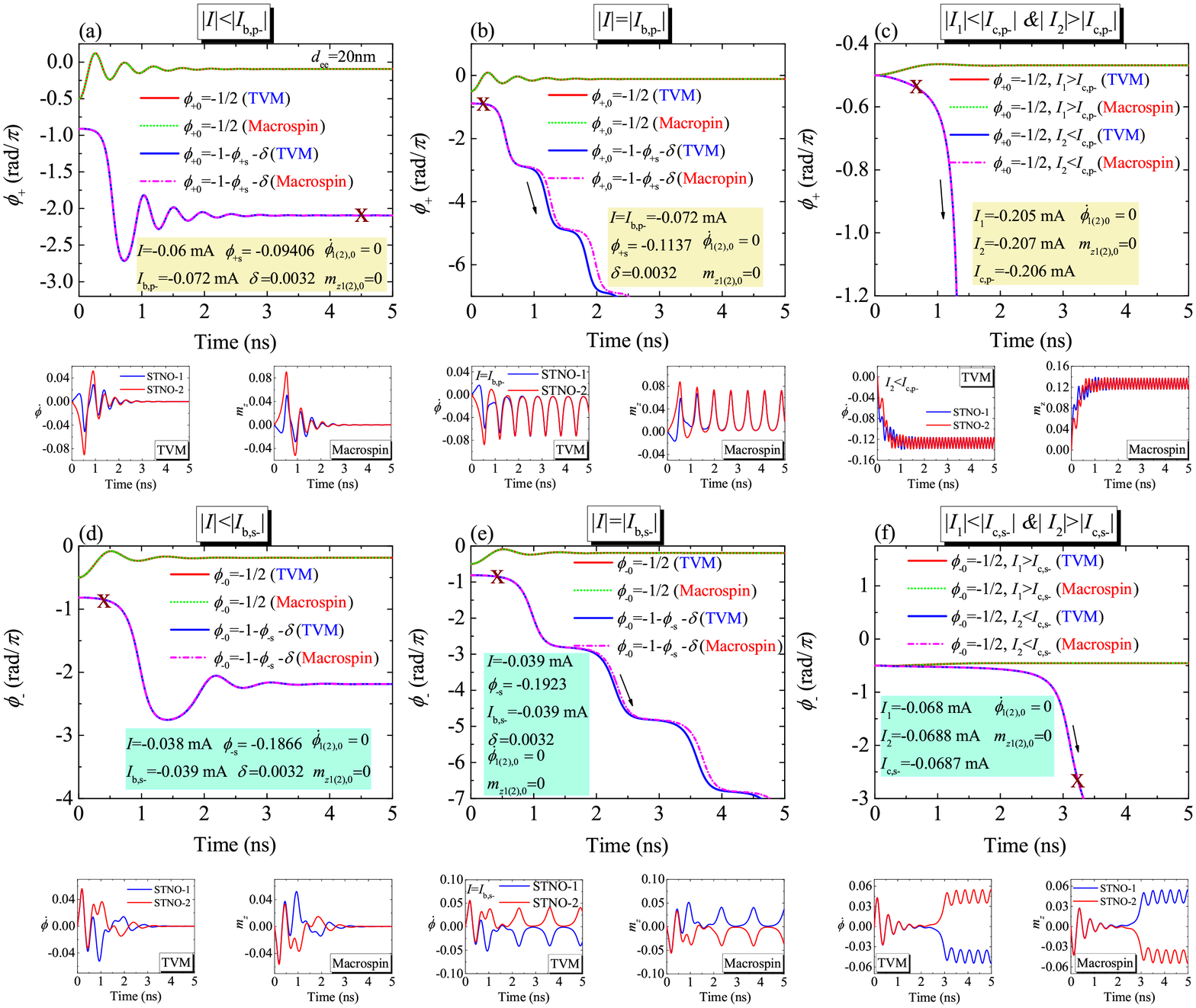}
\end{center}
\caption{(Color Online) Procedures to search for the threshold currents $|I_{\mathrm{b,p(s)}-}|$ and $|I_{\mathrm{c,p(s)}-}|$ of PL states by the TVM and macrospin simulations. Figures (a), (b), and (c) show the time traces of $\phi_{+}$ in parallel connection, and figures (d), (e), and (f) present those of $\phi_{-}$ in serial connection. Also, the separation $d_{\mathrm{ee}}$ are taken to be $20$ nm. In each figure, the red solid curve is the TVM simulation result for the initial state $(\dot{\phi}_{1(2),0}=0,\phi_{+0}=-\pi/2)$, and the green short dot line is the macrospin simulation one for the initial state $(m_{z1(2),0}=0,\phi_{+0}=-\pi/2)$. The blue solid curve is the TVM simulation result after using the slingshot effect, and the mangeta short dash dot line is the macrospin simulation one after performing the same effect, where both of them are marked by the brown sign "X". In the yellow and green boxes are the input parameters including the injected current $I$, the initial velocities $\dot{\phi}_{1(2),0}$ (or $m_{z1(2,0)}$), the initial deviation angle $\delta$ used to prepare the slingshot effect, and the stable phase angles $\phi_{\pm\mathrm{s}}$. The black arrows in Figs. (b),(c), (e), and (f) indicate the driving of the OP states. Below each figure are two panels to display the time traces of $\dot{\phi}_{i}$ and $m_{zi}$, respectively, driven by the slingshot effect. The blue and red solid lines denote the traces of STNO-1 and STNO-2, respectively.}
\label{time_trace_PL}
\end{figure*}

In the following, we would like to generalize the previously mentioned procedures to the cases of trigging PL and AS states in the coupled pair of PERP-STNOs to search for its critical currents $|I_{\mathrm{b,p(s)}}|$, $|I_{\mathrm{c,p(s)}}|$, $|I'_{\mathrm{b,p(s)}}|$, $|I'_{\mathrm{c,p(s)}}|$, and so on. Before doing this, it is observable that the governing equations for the excitations of PL and AS states (see Eq. (\ref{TVMphaslock})), respectively, are similar to a single forced pendulum's governing equation in appearance, indicating that the procedures displayed in Fig. \ref{initialsetting} can also be used to search for the critical currents of these two states.

Therefore, in the case of trigging PL state, see Fig. \ref{time_trace_PL}, the initial states in the very beginning are $(m_{z10}=m_{z20}=0,\phi_{\pm0}=\pm\pi/2)$ or $(\dot{\phi}_{10}=\dot{\phi}_{20}=0,\phi_{\pm0}=\pm\pi/2)$ for the macrospin and TVM models simulations, respectively, where, notably, the signs $\pm$ appearing ahead of $\pi/2$ depend on the positive or negative $\beta'_{\pm}$, respectively, which can be easily seen from the OP precessional directions or $\phi_{\pm\mathrm{s}}$ in direct simulations. Then, after obtaining stable state $(m_{z1(2)}=0,\phi_{\pm\mathrm{s}})$ or $(\dot{\phi}_{1(2)}=0,\phi_{\pm\mathrm{s}})$, $|I_{\mathrm{b,p(s)}}|$ can be find out using the slingshot effect to drive an OP state, see Figs. \ref{time_trace_PL}(a), (b), (d), and (e). Finally, keeping placing the initial state at  $(m_{z10}=m_{z20}=0,\phi_{\pm0}=\pm\pi/2)$ or $(\dot{\phi}_{10}=\dot{\phi}_{20}=0,\phi_{\pm0}=\pm\pi/2)$ and tuning up the current amplitude till the occurrence of an OP state, $|I_{\mathrm{c,p(s)}}|$ can be found out numerically, as shown by Figs. \ref{time_trace_PL} (c) and (f).

\begin{figure*}
\begin{center}
\includegraphics[width=17.7cm]{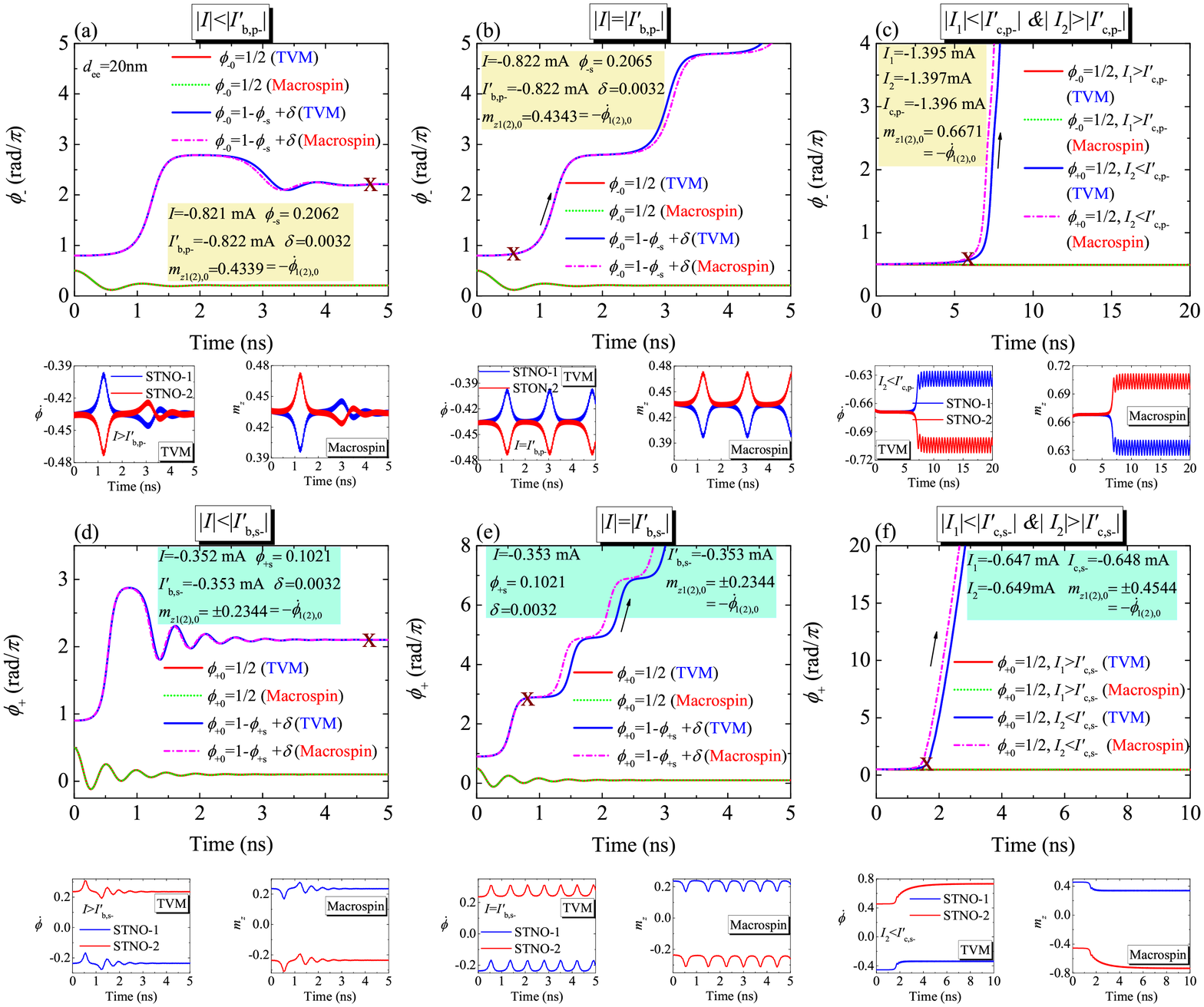}
\end{center}
\caption{(Color Online) The procedures to search for the threshold currents $|I'_{\mathrm{b,p(s)}-}|$ and $|I'_{\mathrm{c,p(s)}-}|$ of AS state by the TVM and macrospin simulations. Figures (a), (b), and (c) show the time traces of $\phi_{-}$ in parallel connection, and figures (d), (e), and (f) present those of $\phi_{+}$ in serial connection. Also, the separation $d_{\mathrm{ee}}$ are taken to be $20$ nm. In each figure, the red solid curve is the TVM simulation result for the initial state $(\dot{\phi}_{1,0}=\pm\dot{\phi}_{2,0}=\dot{\phi}_{0},\phi_{+0}=\pi/2)$, and the green short dot line is the macrospin one for the initial state $(m_{z1,0}=\pm m_{z2,0}=m_{z,0},\phi_{+0}=\pi/2)$, where $\dot{\phi}_{0}=-m_{z,0}\neq0$ and the signs $\pm$ ahead of $\dot{\phi}_{2,0}$ and $m_{z2,0}$ indicate the cases of the parallel and serial connections, respectively. By the way, the colors of these time traces are plotted in the same way as Fig. \ref{time_trace_PL}. }
\label{time_trace_AS}
\end{figure*}

Similarly, in the case of driving AS state, see Fig. \ref{time_trace_AS}, the initial states are $(m_{z10}=\pm m_{z20}= m_{z0},\phi_{\pm0}=\pm\pi/2)$ or $(\dot{\phi}_{10}=\pm\dot{\phi}_{20}=\dot{\phi}_{0},\phi_{\pm0}=\pm\pi/2)$, where $m_{z0}$ and
$\dot{\phi}_{0}$ are given by their phase-locked values. Then, after obtaining the stable state $(m_{z1}=\pm m_{z2}= m_{z0},\phi_{\pm\mathrm{s}})$ or $(\dot{\phi}_{10}=\pm\dot{\phi}_{20}=\dot{\phi}_{0},\phi_{\pm\mathrm{s}})$, $|I'_{\mathrm{b,p(s)}}|$ can be found out using the slingshot effect to drive an AS state, see Figs. \ref{time_trace_AS}(a), (b), (d), and (e). Finally, keeping setting the initial state at $(m_{z10}=\pm m_{z20}= m_{z0},\phi_{\pm0}=\pm\pi/2)$ or $(\dot{\phi}_{10}=\pm\dot{\phi}_{20}=\dot{\phi}_{0},\phi_{\pm0}=\pm\pi/2)$ and adjusting up the current strength till the occurrence of an AS state, $|I'_{\mathrm{c,p(s)}}|$ can be found out in simulations, as shown by Figs. \ref{time_trace_AS} (c) and (f).

Compared with the macrospin, the TVM model simulations, as Figs. \ref{time_trace_PL} and \ref{time_trace_AS} show, have a very good accuracy whatever in transient evolutions or in final stable
states. Besides, the panels apparently reflect that the canonical momentum $p\equiv-m_{z}$, as implied by the Legendre transformation, can be well replaced by the phase angular velocity $\dot{\phi}\approx p$ in the TVM model \cite{Chen2021}. By the way, the time traces of $\dot{\phi}_{i}$ ($m_{zi}$) for the final states shown in both Figs. \ref{time_trace_PL} (b) and (e) (or in both Figs. \ref{time_trace_AS} (b) and (e)), display a comb-like shape, i.e. a non-harmonic oscillation, meaning that the particle pass through the potential's valley at a much faster velocity than it crosses over the potential's peak when the particle's energy gets very close to the energy bottoms of the PL and AS states. If the particle's energy is far from those bottoms, then the time traces of the final states of $\dot{\phi}_{i}$ ($m_{zi}$) will display a harmonic shape instead, as can be seen in Figs. \ref{time_trace_PL} (c) or (f) (or in Figs. \ref{time_trace_AS} (c) or (f)), indicating that the speeds at which the particle passes through the potential's valley and crosses over the potential's peak are comparable to each other.

\subsubsection{\label{D1}Phase Diagrams of Phase-locked State}
\begin{figure*}
\begin{center}
\includegraphics[width=15cm]{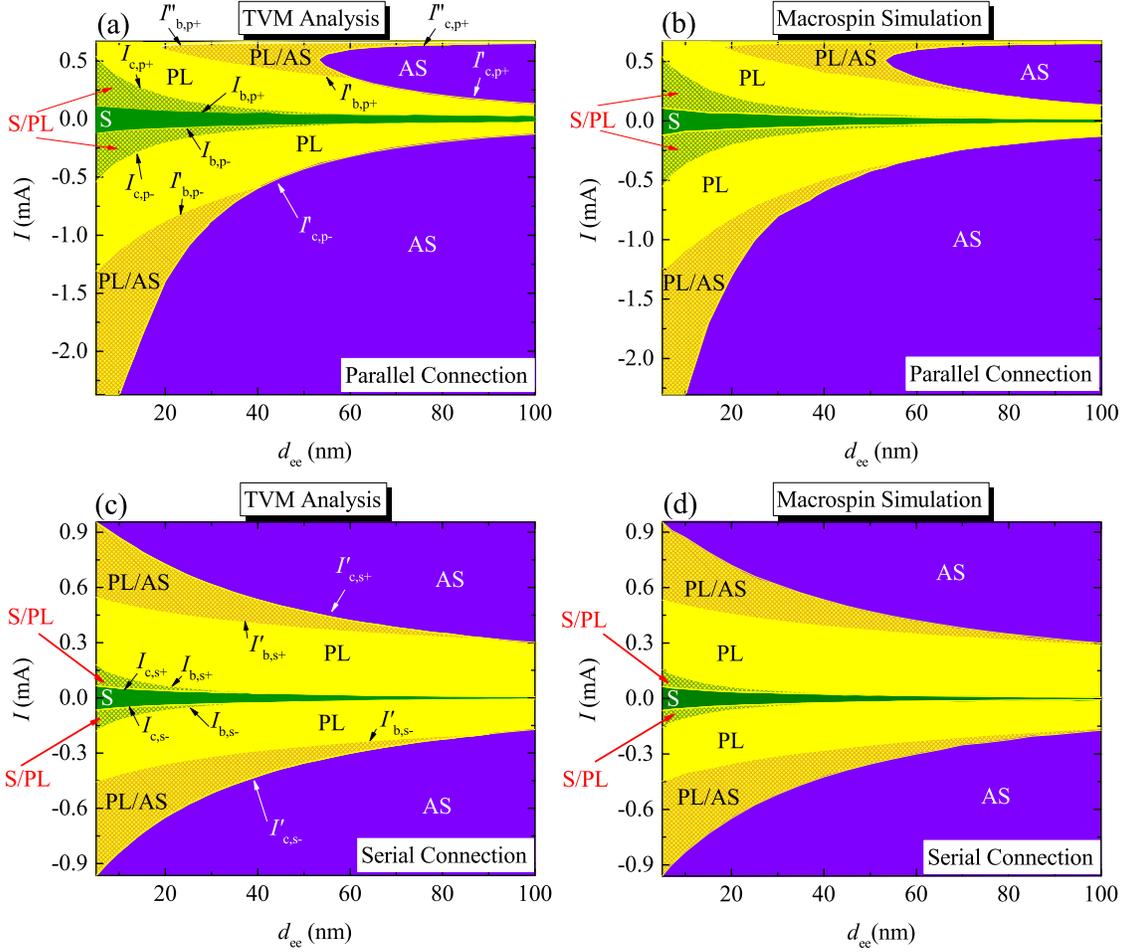}
\end{center}
\caption{(Color online) Phase diagrams for synchronization state as a function of edge-to-edge separation $d_{\mathrm{ee}}$ and current $I$ for the parallel ((a)-(b)) and serial ((c)-(d)) connections, respectively. Diagrams (a) and (c) are calculated from the theoretical model, (b) and (d) are calculated from conducting the macrospin model. Here, the dark cyan, yellow, and purple regions indicate S, PL, and AS states, respectively. Also, the green and yellow areas with yellow and purple dense patterns indicate S/PL and PL/AS states, respectively.  }
\label{phasdia}
\end{figure*}

Having demonstrated how to easily obtain the threshold/critical currents for PL and AS states using the TVM model and macrospin simulations, the dynamic phase diagrams as a function of current $I$ and separation $d_{\mathrm{ee}}$ for the coupled pair of PERP-STNOs in parallel and serial connections, respectively, see Figs. \ref{phasdia} (b) and (d), can be solved easily, which has been done in Ref.\cite{Chen2021}, however, in a not quite accurate way. The reason for that is that we didn't adopt the previously proposed procedure to solve the phase diagram numerically but a more complex and lower efficient one, the hysteretic loop approach, where the current has to be swept forward/backward step by step taking the final stable state given in the previous step as the initial state input in the following step. By the way, there is no need to present the TVM-simulated phase diagrams here because the TVM-analyzed results have been close enough to the macrospin-simulated ones. Thus, we only need to supply the analyzed results to prove the accuracy of the TVM model, as shown by Figs. \ref{phasdia} (a) and (c), using the theoretical techniques developed in section \ref{C}.

As can be seen in Fig. \ref{phasdia}, the analytical results are in very good agreement with those of the macrospin simulations either qualitatively or quantitatively. Also, in Figs. \ref{phasdia} (a) and (b), we have made corrections for some errors existing in Ref. \cite{Chen2021}. More importantly, we would like to stress here that our previous theory termed the generalized pendulum-like model\cite{Chen2021} fails to make predictions for the current range beyond $0.67\mathrm{mA}$ in the serial connection, which, as that work has explained, is due to a big mismatch between the actual phase-locked $m_{z}$ and the free-running states' $m_{z0}$ on which that theory is based. But, here, thanks to Legendre transformation, which completely preserves the dynamical details about $m_{z}$ when developing the TVM model, this difficulty has been pretty well overcome here, as Figs. \ref{phasdia} (c) and (d) display.

Interestingly, in addition to the critical currents $|I'_{\mathrm{c(b),p\pm}}|$ in the parallel case, there are still some other ones like $|I''_{\mathrm{c(b),p+}}|$ existing in the higher positive current range, which are captured by the theoretical analyses and also got the good verification from the simulations, as presented in Figs. \ref{phasdia} (a) and (b).

\subsubsection{ Synchronized and Asynchronized Frequency Responses}
\begin{figure*}
\begin{center}
\includegraphics[width=15cm]{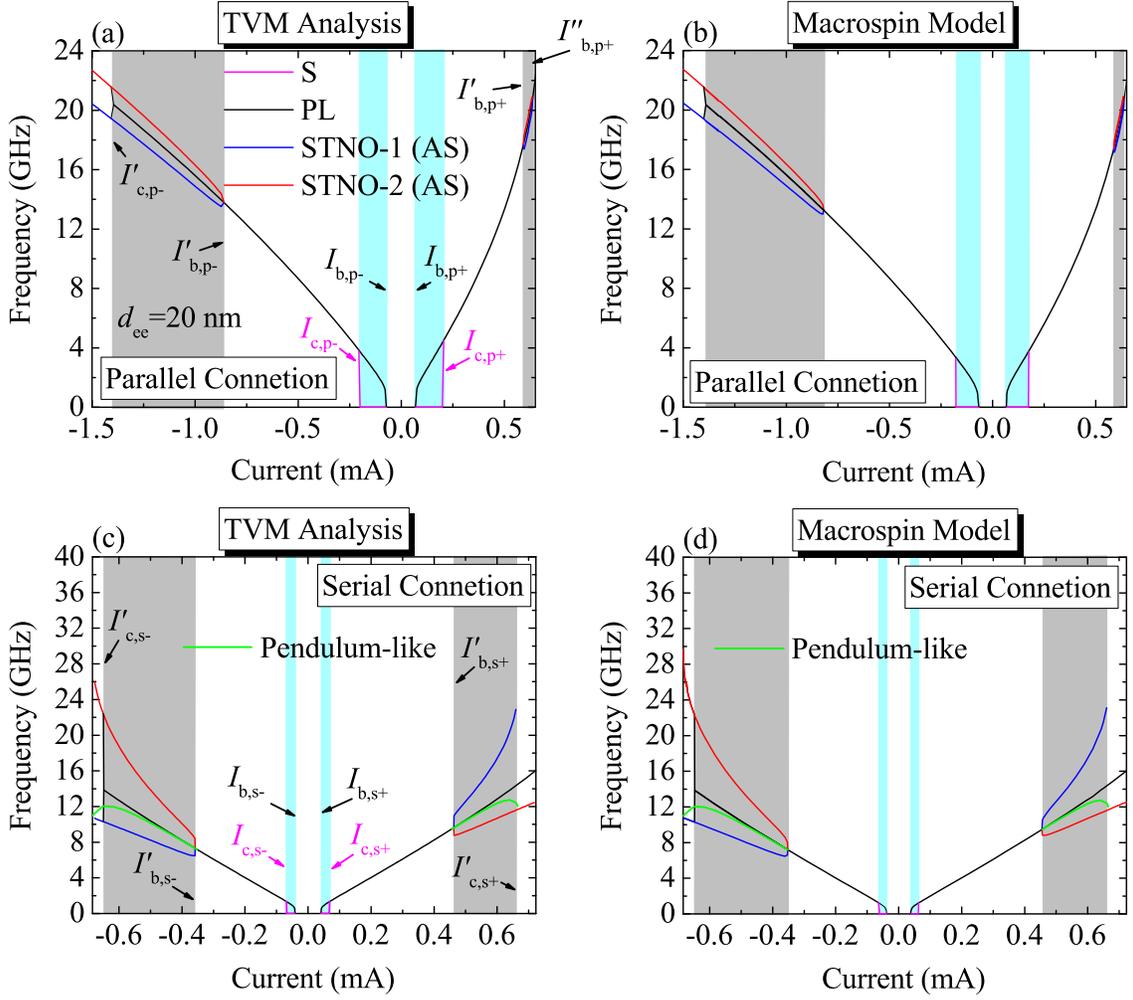}
\end{center}
\caption{(Color online) Synchronized (PL) and asynchronized (AS) frequency responses of phase-locked PERP-STNO pairs against current for the parallel ((a) and (b)) and serial ((c) and (d)) connections, respectively. Here, $d_{\mathrm{ee}}$ and $\alpha$ are taken to be 20 nm and 0.02, respectively. Figures ((a),(c)) and ((b),(d)) denote the results of the analytical and macrospin models, respectively. The coexistent state PL/AS areas are highlighted by the gray color. The coexistent state S/PL areas are highlighted by the light blue color. The magenta, black, blue, and red curves indicate the frequencies of the S, PL, and AS states for STNO-1,2, respectively. The green curves in (c) and (d) indicate the predictions made by the previous theoretical model\cite{Chen2021}.  $I_{\mathrm{b,p(s)\pm}}$ and $I_{\mathrm{c,p(s)\pm}}$ indicate the threshold currents of driving PL state for the parallel and serial connections, respectively. $I'_{\mathrm{b,p(s)\pm}}$, $I''_{\mathrm{b,p+}}$, and $I'_{\mathrm{c,p(s)\pm}}$ are the critical currents of stimulating AS state for these two connections, respectively. }
\label{Hysterfre}
\end{figure*}

Just as implied by the phase diagrams shown in Fig. \ref{phasdia}, the frequency responses as functions of current corresponding to the states of S, S/PL, PL, PL/AS, and AS are all given in Fig. \ref{Hysterfre}, including the TVM-based analytical and macrospin simulation results for the parallel case (see Figs. \ref{Hysterfre} (a) and (b)) and the serial one (see Figs. \ref{Hysterfre} (b) and (d)), respectively, for the purpose of comparison. The qualitative analyses about these frequency responses, especially for the hysteretic frequency responses, have been well done in Ref.\cite{Chen2021}, thus what we like to emphasize here is that compared to our previous theory \cite{Chen2021}, which leads to faulty predictions about the phase-locked frequencies in the current range beyond around $0.57\mathrm{mA}$ in the serial case (see the green curves in Figs. \ref{Hysterfre}(c) and (d)), the TVM model is in very good agrement with the macrospin one in the whole current range instead. This also manifests that the TVM model based on the Legendre transformation is superior to the pendulum-like one based on the expansion at free-running states of STNOs in the completeness of the dynamical information. In the following, this point will be also reflected by the theoretical predictions about the phase-locked angles.

\subsubsection{Phase-locked Phase Angles}
\begin{figure*}
\begin{center}
\includegraphics[width=13.5cm]{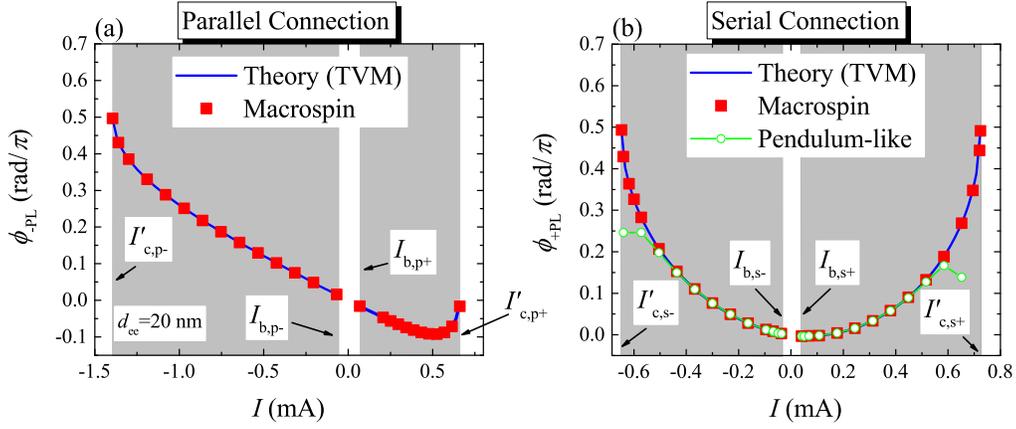}
\end{center}
\caption{(Color online) Stable phase-locked angles $\phi_{\pm\mathrm{PL}}$ as a function of current $I$ for the parallel (a) and serial (b) cases, respectively. Here, $d_{\mathrm{ee}}$ is taken to be 20 nm. The blue solid, green circle, and red square curves indicate the results of the TVM analyses, pendulum-like, and macrospin simulations, respectively. The gray regions mark the PL states.  $I_{\mathrm{b,p(s)}\pm}$ and  $I_{\mathrm{c,p(s)}\pm}$ indicate the threshold currents for the PL states and the critical currents for the AS states, respectively.}
\label{Phasesum}
\end{figure*}

Analyzing the phase-locked angles of a coupled pair of STNOs is very important and meaningful for enhancing the emitted power of STNOs in a synchronized arrangement. Thus, as Fig. \ref{Phasesum} shows, the phase-locked phase angles as functions of current are calculated by the theoretical model, including those of the S/PL, PL, and PL/AS states, which are in very good agreement with the results of the macrospin simulations. The qualitative analyses about these results and the benefits brought by them have received sufficient discussion in Ref. \cite{Chen2021}, thus there is no need to do it again here. But, what's the most important here, as mentioned previously, is that the TVM model perfectly solves the problem of the pendulum-like one failing to correctly deal with the situation where the phase-locked momenta are far from those of the free-running states in the serial case in the higher current range ($|I|\geq0.57 $ mA), as can be seen by the green circle curves in Fig. \ref{Phasesum}(b).

\section{\label{sec:3}Summary and Discussion}
In this work, a terminal velocity motion (TVM) model is developed utilizing Legendre transformation, which is a standard theoretical approach completely transforming the non-linear frequency shift coefficient of an auto-oscillator into the effective mass of a TVM particle. Thus, the dynamical information possessed by the TVM model is nearly as much as that held by the macrospin model, leading to more precise analytical predictions made based on Newton mechanics about the mutual phase-locking of coupled oscillators than those made by our previous model in Ref.\cite{Chen2021}. In addition to improving the drawbacks of the previous model, what can be benefited from the TVM model is that when analyzing the phase-locking of a group of coupled auto-oscillators the phase-locking criterion expressed in configuration space, i.e. a zero frequency or phase angle velocity mismatch between oscillators, is easier to be performed than in phase space, where a non-zero momentum mismatch between oscillators is probably present instead due to an inconsistency in non-linear frequency shift coefficient among oscillators when phase-locking occurs.

Since the TVM model has been proven to be very close to the macrospin one whatever quantitatively or qualitatively in dynamics, we believe using it some issues related to temperature effects on STNOs such as phase noise, generated linewidth, phase slip \cite{Tortarolo2018}, and so forth, could probably be unveiled by this model, helping us to find some effective ways to overcome the shortages caused by temperature. In addition, it is well known that a single PERP-STNO displays some non-uniform states under a higher current density\cite{Ebels2008}, resulting in its generated frequency failing to increase with current, which will hinder its real application about tele-communication.  Therefore, we would like to try to explore the physical mechanism of this phenomenon utilizing several coupled discrete TVM particles by ferromagnetic exchange interactions, and thereby, to find a way to improve this defect.

\section{Data Availability}
The data that support the findings of this study are available from the corresponding author upon reasonable request.

\begin{acknowledgments}
The authors gratefully acknowledge the National Natural Science Foundation of China (Grants No. 61627813 and No. 61571023), the International Collaboration Project No. B16001, the National Key Technology Program of China No. 2017ZX01032101, Beihang Hefei Innovation Research Institute Projects (BHKX-19-01) and (BHKX-19-02) for their financial support of this work.
\end{acknowledgments}

\appendix

\section{\label{appa} General Terminal Velocity model for Non-linear Auto-Oscillatory Systems}
\subsection{\label{appa:1}Legendre Transformation}
In terms of the generalized canonical \textit{cyclic} coordinate\cite{Chen2021}, i.e. \emph{energy-phase angle} representation $(p_{i}\equiv E_{i0},\phi_{i})$ or like \textit{action-angle} representation $(\omega,J)$ introduced in classical mechanics \cite{goldstein2014classical}, the equation of motion for two-dimensional autonomous systems with weak interactions can be written as
\setlength\abovedisplayskip{6pt}
\setlength\belowdisplayskip{6pt}
\begin{eqnarray}
\dot{p_{i}}&=&-\alpha_{i} S_{i}(p_{i})\dot{\phi}_{i}+\beta_{i}(p_{i},\mu_{i})-\frac{\partial H}{\partial\phi_{i}},\nonumber\\
&=&-\alpha_{i} S_{i}(p_{i})\left[\frac{\partial H}{\partial p_{i}}-\frac{\beta_{i}(p_{i},\mu_{i})}{\alpha_{i} S_{i}(p_{i})}\right]-\frac{\partial H}{\partial\phi_{i}},\nonumber\\
\dot{\phi_{i}}&=&\frac{\partial H}{\partial p_{i}},
\label{apppphi}
\end{eqnarray}
where $H(p,\phi)=H_{O}(p)+H_{I}(p,\phi)$ is the total Hamiltonian, $H_{O}(p)=\sum_{i=1}^{n}\int dp'_{i}\dot{\phi_{i}}(p'_{i})=\sum_{i=1}^{n}\int dp'_{i}\left[2\pi/T_{i}(p'_{i})\right]$ is the sum of the Hamiltonian of individual oscillators, and $H_{I}(p,\phi)=(1/2)\sum_{i,j=1(i\neq j)}^{n}U_{I}(p_{i},p_{j},\phi_{i},\phi_{j})$ is the sum of weak interactions with $|H_{I}|\ll|H_{O}|/n$. $T_{i}(p_{i})$ is the period of the dynamic state trajectories $C_{i}(p_{i})$, which can be calculated in the way introduced in Ref. \cite{Chen2021}. Note, here, that if we choose energy as the canonical momentum, then $\dot{\phi}$ must be an angular frequency that is positive instead of an angular velocity. Besides, $S_{i}(p_{i})$ and $\beta_{i}(p_{i},\mu_{i})$ are positive and negative damping functions, respectively, whose definitions are given by Ref. \cite{Chen2021}. Notably, in general,
the energy injected by the non-conservative part during
a single period of the conserved trajectory is negligible compared to the energy level of the trajectory (see Ref.\cite{Chen2021})

In the absence of non-conservative part of Eq. (\ref{apppphi}), i.e. $\alpha_{i}=0$ and $\beta_{i}=0$, Eq. (\ref{apppphi}) can be derived from the variational principle: $\delta I=\delta\int_{t_{1}}^{t_{2}}f(p,\dot{p},\phi,\dot{\phi})dt$, where the integrand is $f(p,\dot{p},\phi,\dot{\phi})=p_{i}\dot{\phi}_{i}-H(p,\phi,t)$. By taking variations of the action $I$ for canonical variables $p$ and $\phi$ independently, one can easily obtain the conservative part of Eq. (\ref{apppphi}).
Notably, due to the existence of \textit{canonical} transformation, the form of the integrand $f$ written down here are not necessarily a Lagrangian (see Ref. \cite{goldstein2014classical}).

Using Legendre transformation, one can replace any one of a set of canonical variables $(p,\phi)$ by the time rate of
its conjugate in Eq. (\ref{apppphi}) as follows: From the conservative part of Eq. (\ref{apppphi}), we have $dH=(\partial H/\partial\phi_{i})d\phi_{i}+(\partial H/\partial p_{i})dp_{i}=-\dot{p}_{i}d\phi_{i}+\dot{\phi}_{i}dp_{i}$. If we want to replace $p_{i}$ with $\dot{\phi}_{i}$, the Legendre transformation can be written to be $L(\dot{\phi},\phi,t)=p_{i}\dot{\phi}_{i}-H(\phi,p,t)$, where $p_{i}=p_{i}(\dot{\phi}_{j},\phi_{j})$ has been used from solving the generalized velocities $\dot{\phi}_{i}=\partial H/\partial p_{i}$, and $L$ is the Lagrangian. Then, we have $dL=(\partial L/\partial\dot{\phi_{i}})d\dot{\phi}_{i}+(\partial L/\partial\phi_{i})d\phi_{i}=p_{i}d\dot{\phi}_{i}+\dot{\phi}_{i}dp_{i}-dH=p_{i}d\dot{\phi}_{i}+\dot{p}_{i}d\phi_{i}$, thereby
 we get a set of equations: $\dot{p}_{i}=\partial L/\partial\phi_{i}$ and $p_{i}=\partial L/\partial\dot
{\phi}_{i}$. Taking a time derivative of the second equation, one obtains the following equations:
\setlength\abovedisplayskip{6pt}
\setlength\belowdisplayskip{6pt}
\begin{eqnarray}
\frac{d}{dt}\left(\frac{\partial L}{\partial\dot{\phi}_{i}}\right)-\frac{\partial L}{\partial\phi_{i}}=0,
\label{configeq}
\end{eqnarray}
which is the equivalent of the conservative part of Eq. (\ref{apppphi}) in configuration space $(\dot{\phi}_{i},\phi_{i})$.

\subsection{\label{appa:2} Lagrangian $L(\dot{\phi},\phi)$}
Owing to the complexity of the form of weak interactions $U_{I}$, the closed form of $p_{i}=p_{i}(\dot{\phi}_{j},\phi_{j})$ is in general difficult to be solved. But, based on the fact that the interaction $H_{I}$ can be taken as a perturbation of $H_{O}$, one can easily obtain $p_{i}=p_{i}(\dot{\phi}_{j},\phi_{j})$ as follows:  from Eq. (\ref{apppphi}), we have
\setlength\abovedisplayskip{6pt}
\setlength\belowdisplayskip{6pt}
\begin{eqnarray}
\dot{\phi}_{i}&=&\frac{\partial H_{O}}{\partial p_{i}}+\frac{\partial H_{I}}{\partial p_{i}},\nonumber\\
&=&\omega_{oi}(p_{i})+Aq_{i}(p_{j},\phi_{j}),
\label{phi}
\end{eqnarray}
where $|Aq_{i}|\ll|\omega_{oi}|$ and $A$ is a very small number measuring the coupling strength. Note that, $\omega_{oi}(p_{i})$ must  be dependent on $p_{i}$ instead of a time constant, implying that the natural frequencies of these oscillators are all non-linear. Also, $q_{i}$ is in general a periodic function of $\phi_{j}$, which means $q_{i}(p_{j},\phi_{j}+2\pi)=q_{i}(p_{j},\phi_{j})$. Then, the momentum $p_{i}$ can be derived from Eq. (\ref{phi}) as
\setlength\abovedisplayskip{6pt}
\setlength\belowdisplayskip{6pt}
\begin{eqnarray}
p_{i}&=&\omega_{oi}^{-1}(\dot{\phi}_{i}-Aq_{i}(p_{j},\phi_{j})),\nonumber\\
&=&\omega_{oi}^{-1}(\dot{\phi}_{i}-Aq_{i}(\omega_{oj}^{-1}(\dot{\phi}_{j}-Aq_{j}(p_{l},\phi_{l})),\phi_{j})),\nonumber\\
&=&\omega_{oi}^{-1}(\dot{\phi}_{i})+\left(\frac{\partial \omega_{oi}^{-1}}{\partial X_{i}}\right)_{\dot{\phi}_{i}}\nonumber\\
&&\times(-A)q_{i}(\omega_{oj}^{-1}(\dot{\phi}_{j}-Aq_{j}(p_{l},\phi_{l})),\phi_{j}))+O(A^{2}),\nonumber\\
&\approx&\omega_{oi}^{-1}(\dot{\phi}_{i})-\left(\frac{\partial \omega_{oi}}{\partial p_{i}}\right)^{-1}_{\omega_{oi}^{-1}(\dot{\phi}_{i})}Aq_{i}(\omega_{oj}^{-1}(\dot{\phi}_{j}),\phi_{j}),
\label{momentump}
\end{eqnarray}
where $X_{i}\equiv\dot{\phi}_{i}-Aq_{i}(p_{j},\phi_{j})$ and $\omega_{oi}(p_{i})$ is monotonous and differentiable. Obviously, replacing $p_{i}$ with $\omega_{oi}^{-1}(\dot{\phi}_{i}-Aq_{i}(p_{j},\phi_{j}))$ on the right-hand side of Eq. (\ref{momentump}) repeatedly, one can easily obtain an approximated expansion of $p_{i}$ at $X_{i}=\dot{\phi}_{i}$ to the first order of $A$, reflecting the fact of $|(\partial\omega_{oi}/\partial p_{i})_{\omega_{oi}^{-1}(\dot{\phi}_{i})}^{-1}Aq_{i}|\ll|\omega^{-1}_{oi}|$. Implementing the inverse operation to Eq. (\ref{momentump}), one can solve $\dot{\phi}_{i}=\dot{\phi}(p_{j},\phi_{j})$ that is exactly the same as Eq. (\ref{phi}).

Using Eq. (\ref{momentump}) and Legendre transformation, the Lagrangian $L(\dot{\phi},\phi)$ can be obtained as follows:
\setlength\abovedisplayskip{6pt}
\setlength\belowdisplayskip{6pt}
\begin{eqnarray}
&L(\dot{\phi},\phi)&=p_{i}\dot{\phi}_{i}-H(p,\phi),\nonumber\\
&\approx&\left[\omega_{oi}^{-1}(\dot{\phi}_{i})-\left(\frac{\partial \omega_{oi}}{\partial p_{i}}\right)^{-1}_{\omega_{oi}^{-1}(\dot{\phi}_{i})}Aq_{i}(\omega_{oj}
^{-1}(\dot{\phi}_{j}),\phi_{j})\right]\dot{\phi}_{i}\nonumber\\
&&-H_{O}\Bigg(\omega_{oi}^{-1}(\dot{\phi}_{i})-\left(\frac{\partial \omega_{oi}}{\partial p_{i}}\right)^{-1}_{\omega_{oi}^{-1}(\dot{\phi}_{i})}\nonumber\\
&&\times Aq_{i}(\omega_{oj}
^{-1}(\dot{\phi}_{j}),\phi_{j})\Bigg)-H_{I}\Bigg(\omega_{oi}^{-1}(\dot{\phi}_{i})\nonumber\\
&&-\left(\frac{\partial \omega_{oi}}{\partial p_{i}}\right)^{-1}_{\omega_{oi}^{-1}(\dot{\phi}_{i})}
Aq_{i}(\omega_{oj}^{-1}(\dot{\phi}_{j}),\phi_{j}),\phi_{i}\Bigg),\nonumber\\
&\approx&\left[\omega_{oi}^{-1}(\dot{\phi}_{i})-\left(\frac{\partial \omega_{oi}}{\partial p_{i}}\right)^{-1}_{\omega_{oi}^{-1}(\dot{\phi}_{i})}Aq_{i}(\omega_{oj}
^{-1}(\dot{\phi}_{j}),\phi_{j})\right]\dot{\phi}_{i}\nonumber\\
&&-H_{O}\left(\omega_{oi}^{-1}(\dot{\phi}_{i})\right)+\dot{\phi}_{i}\left(\frac{\partial \omega_{oi}}{\partial p_{i}}\right)^{-1}_{\omega_{oi}^{-1}(\dot{\phi}_{i})}\nonumber\\
&&\times Aq_{i}(\omega_{oj}^{-1}(\dot{\phi}_{j}),\phi_{j})-H_{I}\left(\omega_{oi}^{-1}(\dot{\phi}_{j}),\phi_{j}\right),
\label{Lagrangian}
\end{eqnarray}
where the Lagrangian has been expanded to the first order of $A$. Note that, although the Lagrangian function here has an approximated form, one can completely return to the original Hamiltonian function by giving it an inverse Legendre transformation.
Subsequently, from $p_{i}=\partial L/\partial\dot{\phi}_{i}$ we get
\setlength\abovedisplayskip{6pt}
\setlength\belowdisplayskip{6pt}
\begin{eqnarray}
&&\frac{d}{dt}\left(\frac{\partial L}{\partial\dot{\phi}_{i}}\right)=\dot{p}_{i},\nonumber\\
&=&\frac{d}{dt}\left[\omega_{oi}^{-1}(\dot{\phi}_{i})-\left(\frac{\partial\omega_{oi}}{\partial p_{i}}\right)^{-1}_{\omega_{oi}^{-1}(\dot{\phi}_{i})}Aq_{i}(\omega_{oj}
^{-1}(\dot{\phi}_{j}),\phi_{j})\right],\nonumber\\
&=&\Bigg[\left(\frac{d\omega_{oi}}{dp_{i}}\right)^{-1}_{\omega_{oi}^{-1}(\dot{\phi}_{i})}\nonumber\\
&&-A\frac{\partial }{\partial\dot{\phi_{i}}}\left[\left(\frac{\partial \omega_{oi}}{\partial p_{i}}\right)^{-1}_{\omega_{oi}^{-1}(\dot{\phi}_{i})}
q_{i}(\omega_{oj}^{-1}(\dot{\phi}_{j}),\phi_{j})\right]\Bigg]\ddot{\phi}_{i}\nonumber\\
&&-A\left(\frac{\partial\omega_{oi}}{\partial p_{i}}\right)^{-1}_{\omega_{oi}^{-1}(\dot{\phi}_{i})}\sum_{l=1(l\neq i)}^{n}\left[\frac{\partial}{\partial\dot{\phi_{l}}}q_{i}(\omega_{oj}
^{-1}(\dot{\phi}_{j}),\phi_{j})
\right]\ddot{\phi}_{l}\nonumber\\
&&-A\left(\frac{\partial\omega_{oi}}{\partial p_{i}}\right)^{-1}_{\omega_{oi}^{-1}(\dot{\phi}_{i})}\sum_{l=1}^{n}\left[\frac{\partial}{\partial\phi_{l}}q_{i}(\omega_{oj}
^{-1}(\dot{\phi}_{j}),\phi_{j})\right]
\dot{\phi}_{l},\nonumber
\end{eqnarray}
and
\setlength\abovedisplayskip{6pt}
\setlength\belowdisplayskip{6pt}
\begin{eqnarray}
\frac{\partial L}{\partial\phi_{i}}&=&-\frac{\partial H_{I}}{\partial\phi_{i}}.\nonumber
\end{eqnarray}
Then, Eq. (\ref{configeq}) becomes
\setlength\abovedisplayskip{6pt}
\setlength\belowdisplayskip{6pt}
\begin{eqnarray}
m_{\mathrm{eff},i}(\omega_{oj}^{-1}(\dot{\phi}_{j}),\phi_{j})\ddot{\phi}_{i}&=&-\frac{\partial H_{I}}{\partial\phi_{i}}+A\left(\frac{\partial\omega_{oi}}{\partial p_{i}}\right)^{-1}_{\omega_{oi}^{-1}(\dot{\phi}_{i})}\nonumber\\
&&\times\left[\sum_{l=1(l\neq i)}^{n}\frac{\partial q_{i}}{\partial\dot{\phi}_{l}}\ddot{\phi}_{l}+\sum_{l=1}^{n}\frac{\partial q_{i}}{\partial\phi_{l}}\dot{\phi}_{l}\right].\nonumber\\
\label{conservedeq}
\end{eqnarray}
where the effective mass of inertia is
\setlength\abovedisplayskip{6pt}
\setlength\belowdisplayskip{6pt}
\begin{eqnarray}
m_{\mathrm{eff},i}(\omega_{oj}^{-1}(\dot{\phi}_{j}),\phi_{j})&\equiv&\left(\frac{d\omega_{oi}}{dp_{i}}\right)
_{\omega^{-1}_{oi}(\dot{\phi}_{i})}^{-1}\nonumber\\
&&-A\frac{\partial}{\partial\dot{\phi}_{i}}\left[\left(\frac{d\omega_{oi}}{dp_{i}}\right)
_{\omega^{-1}_{oi}(\dot{\phi}_{i})}^{-1}q_{i}\right].\nonumber\\
\label{effmass}
\end{eqnarray}
Apparently, the second term in the effective mass is the modification from the interactions, which in general can be neglected due to the smallness of $A$. So, $m_{\mathrm{eff},i}(\omega_{oi}^{-1}(\dot{\phi}_{i}))\approx(d\omega_{oi}/dp_{i})_{\omega_{oi}^{-1}(\dot{\phi}_{i})}^{-1}$  is related to the non-linear frequency shift coefficient $d\omega_{oi}/dp_{i}$.

Since the term related to $\ddot{\phi}_{l}$ is proportional to $A$, it can be further neglected in Eq. (\ref{conservedeq}). For coupled particles with strong non-linear frequency shift coefficients $d\omega_{oi}/dp_{i}$, where the last term appearing in Eq. (\ref{conservedeq}) can be neglected, one can consider a certain scalar quantity defined as $W(\dot{\phi},\phi)=\sum_{i}W_{oi}(\dot{\phi}_{i})+H_{I}(\omega_{oj}^{-1}(\dot{\phi}_{j}),\phi_{j})$, in which $W_{oi}(\dot{\phi}_{i})=\int^{\dot{\phi}_{i}} d\dot{\phi}'_{i}\left[m_{\mathrm{eff},i}(\omega_{oi}^{-1}(\dot{\phi}'_{i}))\dot{\phi}'_{i}\right]$. The
 balance equation for this quantity is given as follows using Eq. (\ref{conservedeq}):
\setlength\abovedisplayskip{6pt}
\setlength\belowdisplayskip{6pt}
\begin{eqnarray}
\frac{dW}{dt}&=&\sum_{i}^{n}\left[m_{\mathrm{eff},i}(\omega_{oj}^{-1}(\dot{\phi}_{j}))\ddot{\phi}_{i}
+\frac{\partial H_{I}}{\partial\phi_{i}}\right]\dot{\phi}_{i}+\frac{\partial H_{I}}{\partial\dot{\phi}_{i}}\ddot{\phi}_{i},\nonumber\\
&\approx&\sum_{i}^{n}\frac{\partial H_{I}}{\partial\dot{\phi}_{i}}\ddot{\phi}_{i}\propto A^{2},
\label{}
\end{eqnarray}
 reflecting that this quantity, strictly speaking, is not conserved. But, thanks to the sufficient smallness of $A$ and the time-periodicity of the last term in $dW/dt$, leading to $\langle\Delta W\rangle_{t}=(1/t)\int_{0}^{t}dt'\dot{W}=(1/t)\int_{0}^{t}dt'\left(\partial H_{I}/\partial\dot{\phi}_{i}\right)\ddot{\phi}_{i}=0$ where the chosen time interval $t$ should be much longer than the time-period $T$ of the system, it can be taken to be nearly conserved here, which is called \textit{quasi-energy conservation}.

\subsection{\label{appa:3} Dissipation Effect: $\alpha_{i}\neq0$, $\beta_{i}\neq0$}
In order to express Eq. (\ref{apppphi}) completely in configuration space, we have to put a \textit{dissipative} force $F_{\mathrm{d},i}$ into Eq. (\ref{conservedeq}) using the exact Hamiltonian balance equation:
\setlength\abovedisplayskip{6pt}
\setlength\belowdisplayskip{6pt}
\begin{eqnarray}
\frac{dH}{dt}&=&\sum_{i=1}^{n}\frac{\partial H}{\partial\phi_{i}}\dot{\phi}_{i}+\frac{\partial H}{\partial p_{i}}\dot{p}_{i},\nonumber\\
&\approx&\sum_{i=1}^{n}F_{\mathrm{d},i}\dot{\phi}_{i},\nonumber
\end{eqnarray}
where the term related to $\dot{p}_{i}$ has bee dropped out due to $|\dot{p}_{i}|\sim|-\partial H/\partial\phi_{i}|\ll|\dot{\phi}_{i}|\sim|\partial H/\partial p_{i}|$.
So, we get the dissipative force as $F_{\mathrm{d},i}=-\partial f_{\mathrm{dis}}/\partial\dot{\phi}_{i}$ and the \textit{dissipation}
function $f_{\mathrm{dis}}=(1/2)\alpha_{i}S_{i}(p_{i})\dot{\phi}_{i}^{2}-\beta_{i}(p_{i},\mu_{i})\dot{\phi}_{i}$.
Finally, from Eq. (\ref{configeq}) with dissipation, i.e. $d(\partial L/\partial\dot{\phi}_{i})/dt-\partial L/\partial\phi_{i}
=-\partial f_{\mathrm{dis}}/\partial\dot{\phi}_{i}$, we obtain the equivalent of Eq. (\ref{apppphi}) in configuration space:
\setlength\abovedisplayskip{6pt}
\setlength\belowdisplayskip{6pt}
\begin{eqnarray}
\ddot{\phi}_{i}&=&\left[\frac{1}{m_{\mathrm{eff},i}(\omega_{oi}^{-1}(\dot{\phi}_{i}))}\right]
\Bigg[-\alpha_{i}S_{i}(\omega_{oi}^{-1}(\dot{\phi}_{i}))\dot{\phi}_{i}\nonumber\\
&&+\beta_{i}(\omega_{oi}^{-1}(\dot{\phi}_{i}),\mu_{i})-\frac{\partial H_{I}}{\partial\phi_{i}}\Bigg]+\frac{A}{m_{\mathrm{eff},i}(\omega_{oi}^{-1}(\dot{\phi}_{i}))}\nonumber\\
&&\times\left(\frac{\partial m_{\mathrm{eff},i}}{\partial\dot{\phi}_{i}}\right) q_{i}\ddot{\phi}_{i}+A\sum_{l=1}^{n}\left(\frac{\partial q_{i}}{\partial\dot{\phi}_{l}}\ddot{\phi}_{l}+\frac{\partial q_{i}}{\partial\phi_{l}}\dot{\phi}_{l}\right),\nonumber\\
\label{TVM}
\end{eqnarray}
 where we have used the approximations $\alpha_{i}S(p_{i})\approx\alpha_{i}S_{i}(\omega_{oi}^{-1}(\dot{\phi}_{i}))$ and $\beta_{i}(p_{i},\mu_{i})\approx\beta_{i}(\omega_{oi}^{-1}(\dot{\phi}_{i}),\mu_{i})$ since $\alpha_{i}A$ and $|(\partial\beta_{i}/\partial p_{i})_{p_{i}=\omega_{oi}^{-1}(\dot{\phi}_{i})}|A$ are both much smaller than $\alpha_{i}$ and $|\beta_{i}|$ to be neglected.

\subsection{\label{appa:4}Generalized TVM Model for a Quasi-linear Auto-oscillator: Adler Equation}
For a \textit{quasi-linear} auto oscillator having an extremely small non-linear frequency shift coefficient $|d\omega/dp|\ll1$, or equivalently, having an extremely large effective mass $|m_{\mathrm{eff}}(\dot{\phi})|\gg1$, it is very hard to
give rise to a significant change of such an oscillator's frequency through positive/negative damping
effects or external interactions within its single period $T=2\pi/\omega$.
Then, under $|m_{\mathrm{eff}}(\dot{\phi})|\gg1$, Eq. (\ref{TVM}) can be reduced to be
\setlength\abovedisplayskip{6pt}
\setlength\belowdisplayskip{6pt}
\begin{eqnarray}
\ddot{\phi}_{i}&\approx&A\frac{dq_{i}}{dt}.\nonumber
\label{}
\end{eqnarray}
Moreover, integrating time on both sides of the above equation, one gets the first time derivative equation of $\phi_{i}$
\setlength\abovedisplayskip{6pt}
\setlength\belowdisplayskip{6pt}
\begin{eqnarray}
\dot{\phi}_{i}&=&\omega_{i0}+Aq_{i}\left(\omega^{-1}_{oj}(\dot{\phi}_{j}),\phi_{j}\right),\nonumber\\
&=&\omega_{i0}+\left(\frac{\partial H_{I}}{\partial p_{i}}\right)_{p_{j}=\omega^{-1}_{oj}(\dot{\phi}_{j})},
\label{Adlerequa}
\end{eqnarray}
where the frequency $\omega_{i0}$ is a constant of integration. As a result, we have proven that a non-linear auto-oscillator described by a generalized TVM model (Eq. (\ref{TVM})) can cover the case for \textit{quasi-linear} auto-oscillators described by Eq. (\ref{Adlerequa}), i.e. Adler equation.

\subsection{\label{appa:5}Generalized TVM Model for a coupled pair of PERP-STNOs}
The dynamics of PERP-STNOs can be assumed to be governed by the Landau-Lifshitz-Gilbert-Slonczewski (LLGS)
equation with the STT effect\cite{HaoHsuan2016,HaoHsuan2018,Chen2021}, i.e. the macrospin model:
\begin{eqnarray}
\frac{d \mathbf{m}_{i}}{d\tau}&=&-\left(\nabla_{\mathrm{m}_{i}} E\right) \times \mathbf{m}_{i}+\alpha \mathbf{m}_{i} \times\left(\frac{d \mathbf{m}_{i}}{d \tau}\right)\nonumber\\
&&+a_{J_{i}}\left(m_{iz}\right)\left[\mathbf{m}_{i} \times\left(\mathbf{m}_{i} \times \mathbf{e}_{z}\right)\right],
\label{LLGinLab}
\end{eqnarray}
where $\textbf{m}_{i} = \textbf{M}_{i}/M_{s}$  is the unit vector of the free layer magnetizations and  $M_{s}$ is the saturation magnetization. $\tau$ is the scaled time $\tau=(4\pi M_{s}\gamma)t$, where $ \gamma= 1.76\times10^{7}$ $\textrm{Oe}^{-1}\cdot \textrm{s}^{-1}$ is the gyromagnetic ratio. $\alpha$ is the Gilbert damping constant.
 Compared to Eq. (\ref{vectformappro}), the vectors $\mathbf{x}$ and $\mathbf{e}_{n}$ have been both replaced by the magnetization unit vector $\mathbf{m}$ in the LLGS equation.

 The third term on the right-hand side of Eq. (\ref{LLGinLab}) is the STT term, where $\mathbf{e}_{p}=\mathbf{e}_{z}$ is the unit vector of magnetization of the PL layer. $a_{Ji}(\textbf{m}_{i})=A_{Ji}(\textbf{m}_{i})(4\pi M_{s}\gamma)^{-1}=a_{Ji0}\varepsilon_{i}(m_{iz},P_{i},\Lambda_{i})$ is the scaled STT strength, and $a_{Ji0}=(\hbar J_{i}/8\pi eM_{s}^{2}d)$. Here, $J$ is the injected current density flowing through the STNO, $d$ is the free layer thickness, and $m_{z}$ is the projection of the FL magnetization unit vector on $\mathbf{e}_{p}$. In addition, $\varepsilon_{i}(m_{iz},P_{i},\Lambda_{i})=P_{i}\Lambda_{i}^{2}/[(\Lambda_{i}^{2}+1)
+(\Lambda_{i}^{2}-1)m_{iz}]$ is the asymmetric factor of the Slonczewski STT\cite{XiaoJiang2004}, where $P_{i}$ and $\Lambda_{i}$ are dimensionless quantities giving the spin-polarization efficiencies.
In our study, the lateral dimension of FL is supposed to be $60\times60$ $\textrm {nm}^{2}$ and the thickness $d=3$ $\textrm{nm}$. The standard material parameters of Permalloy $(\textrm{Ni}_{80}\textrm{Fe}_{20})$ are used for the FL: saturation magnetization $M_{s}=866$ $\textrm{emu}/\textrm{cm}^{3}$, and dimensionless quantities of spin-polarization efficiency $P=0.38$, and $\Lambda=1.8$\cite{XiaoJiang2004}.\\

Introduced the background details about an individual PERP-STNO, in the following we would like to derive the equivalent of Eq.(\ref{LLGinLab}), i.e. Eq.(\ref{TVM}) for a coupled PERP-STNO pairs in the configuration space $(\dot{\phi}_{i},\phi_{i})$. Due to the axial symmetry of an individual PERP-STNO, including the demagnetization energy as well as the spin polarization vector, one can choose the canonical momentum $p_{i}\equiv-m_{iz}$. Thereby, in this model, the total Hamiltonian ($H\equiv E$) of a pair of PERP-STNOs coupled by a dipolar interaction reads
\setlength\abovedisplayskip{6pt}
\setlength\belowdisplayskip{6pt}
\begin{eqnarray}
H(p,\phi)&=&\sum_{i=1}^{2}H_{Oi}(p_{i})+H_{I}(p_{1},p_{2},\phi_{1},\phi_{2}),\nonumber\\
&=&\frac{k}{2}\sum_{i=1}^{2}p_{i}^{2}-A_{\mathrm{disc}}(d_{\mathrm{ee}})\Bigg\{\frac{1}{2}\sqrt{(1-p_{1}^{2})(1-p_{2}^{2})}\nonumber\\
&&\times\left[3\cos(\phi_{1}+\phi_{2})+\cos(\phi_{1}-\phi_{2})\right]-p_{1}p_{2}\Bigg\},\nonumber\\
\label{HamitonPERPSTNOs}
\end{eqnarray}
where the momentum $p_{i}\equiv-m_{iz}$. Note that, when $d_{ee}$ ranges from $5$ nm to $100$ nm, the value of $A_{\mathrm{disc}}(d_{ee})$ ranges from $1.72\times10^{-4}$ to $4.75\times10^{-3}$, which is vary small compared with that of the demagnetization field strength $|k|=1$. Note that, in the viewpoint of the micromagnetic model, the actual value of $k$ must be slightly smaller than that in the macrospin model, where the reasonable ratio of the micromagnetic model to the macrospin one should be around $0.7\sim0.8$ for the size of the free layer $30\times 30\times3 \mathrm{nm}^{3}$.

From Eq. (\ref{phi}), one gets
\setlength\abovedisplayskip{6pt}
\setlength\belowdisplayskip{6pt}
\begin{eqnarray}
\dot{\phi}_{i}&=&kp_{i}+\frac{\partial H_{I}}{\partial p_{i}}\hspace{1cm} (i,j=1,2), \nonumber\\
&=&kp_{i}-A_{\mathrm{disc}}(d_{\mathrm{ee}})\Bigg\{-\frac{1}{2}p_{i}\sqrt{\frac{1-p_{j}^{2}}{1-p_{i}^{2}}}\big[3\cos(\phi_{1}
+\phi_{2})\nonumber\\
&&+\cos(\phi_{1}-\phi_{2})\big]-p_{j}\Bigg\}. \nonumber\\
\label{dotphiasafuncp}
\end{eqnarray}
where $i,j=1,2$. Then, according to Eq. (\ref{momentump}), the canonical momentum $p_{i}$ can be expressed as
\setlength\abovedisplayskip{6pt}
\setlength\belowdisplayskip{6pt}
\begin{eqnarray}
p_{i}&=&\dot{\phi}_{i}'+A'_{\mathrm{disc}}(d_{\mathrm{ee}})
\Bigg\{-\frac{1}{2}\dot{\phi}_{i}'\sqrt{\frac{1-\dot{\phi}_{j}'^{2}}
{1-\dot{\phi}_{i}'^{2}}}\nonumber\\
&&\times[3\cos(\phi_{1}+\phi_{2})+\cos(\phi_{1}-\phi_{2})]-\dot{\phi}_{j}'\Bigg\}, \nonumber\\
\label{pasafuncdotphi}
\end{eqnarray}
where $\dot{\phi}_{i}'\equiv\dot{\phi}_{i}/k$, $A'_{\mathrm{disc}}(d_{\mathrm{ee}})\equiv A_{\mathrm{disc}}(d_{\mathrm{ee}})/k$, and $(d\omega_{oi}/dp_{i})_{\omega_{oi}^{-1}(\dot{\phi}_{i})}^{-1}=k^{-1}$. Thus, the effective mass here is $m_{\mathrm{eff},i}(\omega_{oi}^{-1}(\dot{\phi}_{i}))=k^{-1}$.

In the coordinate system $(p_{i},\phi_{i})$, Eq.(\ref{LLGinLab}) can be rewritten to be
\setlength\abovedisplayskip{6pt}
\setlength\belowdisplayskip{6pt}
\begin{eqnarray}
\dot{p}_{i}&=&-\left(\frac{\alpha_{i}}{1+\alpha_{i}^{2}}\right)\left(1-p_{i}^{2}\right)\left[\frac{\partial H}{\partial p_{i}}-\frac{a_{Ji}(-p_{i})}{\alpha_{i}}\right]\nonumber\\
&&-\left(\frac{1}{1+\alpha_{i}^{2}}\right)\frac{\partial H}{\partial\phi_{i}},\nonumber\\
\dot{\phi}_{i}&=&\left(\frac{1}{1+\alpha_{i}^{2}}\right)\frac{\partial H}{\partial p_{i}}-\left(\frac{\alpha_{i}}{1+\alpha_{i}^{2}}\right)
\left(\frac{1}{1-p_{i}^{2}}\right)\frac{\partial H}{\partial\phi_{i}}\nonumber\\
&&+\left(\frac{\alpha_{i}}{1+\alpha_{i}^{2}}\right)a_{Ji}(-p_{i}).
\label{LLGSinclyn}
\end{eqnarray}
Then, using Eq. (\ref{LLGSinclyn}) the exact Hamiltonian balance equation is
\setlength\abovedisplayskip{6pt}
\setlength\belowdisplayskip{6pt}
\begin{eqnarray}
\frac{dH}{dt}&=&\sum_{i=1}^{2}\frac{\partial H}{\partial\phi_{i}}\dot{\phi}_{i}+\frac{\partial H}{\partial p_{i}}\dot{p}_{i},\nonumber\\
&=&\sum_{i=1}^{2}\Bigg\{\frac{\partial H}{\partial p_{i}}\bigg[-\left(\frac{\alpha_{i}}{1+\alpha_{i}^{2}}\right)(1-p_{i}^{2})\nonumber\\
&&\times\left(\frac{\partial H}{\partial p_{i}}-\frac{a_{Ji}(-p_{i})}{\alpha_{i}}\right)\bigg]-\left(\frac{\alpha_{i}}{1+\alpha_{i}^{2}}\right)\frac{1}{1-p_{i}^{2}}\nonumber\\
&&\times\left(\frac{\partial H}{\partial\phi_{i}}\right)^{2}+\left(\frac{\alpha_{i}}{1+\alpha_{i}^{2}}\right)a_{Ji}(-p_{i})\frac{\partial H}{\partial\phi_{i}}\Bigg\},\nonumber\\
&\approx&\sum_{i=1}^{2}\Bigg\{-\alpha_{i}(1+\alpha_{i}^{2})(1-p_{i}^{2})\dot{\phi}_{i}^{2}+\left[\frac{a_{Ji}(-p_{i})}
{1+\alpha_{i}^{2}}\right]\nonumber\\
&&\times(1-p_{i}^{2})\dot{\phi}_{i}\Bigg\},\nonumber\\
&\approx&\sum_{i=1}^{2}-\alpha_{i}(1-p_{i}^{2})\dot{\phi}_{i}^{2}+a_{Ji}(-p_{i})(1-p_{i}^{2})\dot{\phi}_{i},\nonumber\\
&=&\sum_{i=1}^{2}F_{\mathrm{d},i}(p_{i})\dot{\phi}_{i},\nonumber
\end{eqnarray}
where $F_{\mathrm{d},i}(p_{i})=-\alpha_{i}(1-p_{i}^{2})\dot{\phi}_{i}+a_{Ji}(-p_{i})(1-p_{i}^{2})$, where $\partial H/\partial\phi_{i}=\partial H_{I}/\partial\phi_{i}=Aq(p_{i},\phi_{i})\propto A(1-p_{i}^{2})^{N/2}$ and $N$ is a positive
integer, and where the terms related to $\alpha_{i}^{n}A^{m}(n,m=1,2,3)$, $\alpha_{i}^{n}|a_{Ji}|A(n=1,2,3)$ and $\alpha_{i}^{n}|a_{Ji}|^{m}(n,m=1,2,3)$ are extremely small to be neglected.
According to Eq. (\ref{LLGinLab}) the positive damping factors $S_{i}(\omega_{oi}^{-1}(\dot{\phi}_{i}))$ and negative damping ones $\beta_{i}(\omega_{oi}^{-1}(\dot{\phi}_{i}),\mu_{i})$ are given as follows:
\setlength\abovedisplayskip{6pt}
\setlength\belowdisplayskip{6pt}
\begin{eqnarray}
S_{i}(\omega_{oi}^{-1}(\dot{\phi}_{i}))&=&1-\dot{\phi}_{i}'^{2},\nonumber\\
\beta_{i}(\omega_{oi}^{-1}(\dot{\phi}_{i})
,\mu_{i})&=&\left(1-\dot{\phi}_{i}'^{2}\right)\frac{a_{Ji0}P_{i}\Lambda_{i}^{2}}{(\Lambda_{i}^{2}+1)
+(\Lambda_{i}^{2}-1)\left(-\dot{\phi}_{i}'\right)}.\nonumber
\label{}
\end{eqnarray}
 Moreover, we have
\setlength\abovedisplayskip{6pt}
\setlength\belowdisplayskip{6pt}
\begin{eqnarray}
-\frac{\partial H_{I}}{\partial\phi_{1}}&=&-\frac{A_{\mathrm{disc}}(d_{\mathrm{ee}})}{2}\sqrt{\left(1-\dot{\phi}_{1}'^{2}\right)
\left(1-\dot{\phi}_{2}'^{2}\right)}\nonumber\\
&&\times\left[3\sin(\phi_{1}+\phi_{2})+\sin(\phi_{1}-\phi_{2})\right],\nonumber\\
-\frac{\partial H_{I}}{\partial\phi_{2}}&=&-\frac{A_{\mathrm{disc}}(d_{\mathrm{ee}})}{2}\sqrt{\left(1-\dot{\phi}_{1}'^{2}\right)
\left(1-\dot{\phi}_{2}'^{2}\right)}\nonumber\\
&&\times\left[3\sin(\phi_{1}+\phi_{2})-\sin(\phi_{1}-\phi_{2})\right],\nonumber
\label{}
\end{eqnarray}
and
\setlength\abovedisplayskip{6pt}
\setlength\belowdisplayskip{6pt}
\begin{eqnarray}
A\sum_{l=1}^{n}\frac{\partial q_{1}}{\partial\phi_{l}}\dot{\phi}_{l}&=&A_{\mathrm{disc}}(d_{\mathrm{ee}})\Bigg[\left(\dot{\phi}_{1}+\dot{\phi}_{2}\right)
\left(\frac{3\dot{\phi}_{1}'}{2}
\sqrt{\frac{1-\dot{\phi}_{2}'^{2}}{1-\dot{\phi}_{1}'^{2}}}\right)\nonumber\\
&&\times\sin(\phi_{1}+\phi_{2})+\left(\dot{\phi}_{1}-\dot{\phi}_{2}\right)\nonumber\\
&&\left(\frac{\dot{\phi}_{1}'}{2}
\sqrt{\frac{1-\dot{\phi}_{2}'^{2}}{1-\dot{\phi}_{1}'^{2}}}\right)\times\sin(\phi_{1}-\phi_{2})\Bigg],\nonumber\\
A\sum_{l=1}^{n}\frac{\partial
 q_{2}}{\partial\phi_{l}}\dot{\phi}_{l}&=&A_{\mathrm{disc}}(d_{\mathrm{ee}})\Bigg[\left(\dot{\phi}_{1}+\dot{\phi}_{2}\right)
\left(\frac{3\dot{\phi}_{2}'}{2}
\sqrt{\frac{1-\dot{\phi}_{1}'^{2}}{1-\dot{\phi}_{2}'^{2}}}\right)\nonumber\\
&&\times\sin(\phi_{1}+\phi_{2})+\left(\dot{\phi}_{1}-\dot{\phi}_{2}\right)\nonumber\\
&&\left(\frac{\dot{\phi}_{2}'}{2}
\sqrt{\frac{1-\dot{\phi}_{1}'^{2}}{1-\dot{\phi}_{2}'^{2}}}\right)\times\sin(\phi_{1}-\phi_{2})\Bigg].\nonumber
\label{}
\end{eqnarray}
Due to $\partial m_{\mathrm{eff},i}/\partial\dot{\phi}_{i}=0$ and $\ddot{\phi}_{l}\propto A_{\mathrm{disc}}(d_{\mathrm{ee}})$, the third and fourth terms on the right-hand side of Eq. (\ref{TVM}) can be reasonably neglected. Finally, from Eq. (\ref{TVM}), we obtain a set of equations of motion for a pair of coupled PERP-STNOs in configuration space.

\nocite{*}

\bibliography{HH_phaselocking}

\end{document}